\documentclass[iop,apjl,tighten]{emulateapj}

\usepackage{hyperref}
\usepackage{mathptmx}

\shorttitle{SOLAR SIBLING CANDIDATES}
\shortauthors{RAM\'IREZ ET AL.}
\submitted{To appear in the Astrophysical Journal}

\newcommand{\feh}{\mathrm{[Fe/H]}}
\newcommand{\teff}{T_\mathrm{eff}}
\newcommand{\logg}{\log g}
\newcommand{\vt}{v_t}

\newcommand{\fei}{Fe\,\textsc{i}}
\newcommand{\feii}{Fe\,\textsc{ii}}
\newcommand{\kms}{km\,s$^{-1}$}

\begin{document}

\title{ELEMENTAL ABUNDANCES OF SOLAR SIBLING CANDIDATES}

\author{I.\,Ram\'irez\altaffilmark{1},
        A.\,T.\,Bajkova\altaffilmark{2},
        V.\,V.\,Bobylev\altaffilmark{2,3},
        I.\,U.\,Roederer\altaffilmark{4},
        D.\,L.\,Lambert\altaffilmark{1},\\
        M.\,Endl\altaffilmark{1},
        W.\,D.\,Cochran\altaffilmark{1},
        P.\,J.\,MacQueen\altaffilmark{1}, and
        R.\,A.\,Wittenmyer\altaffilmark{5,6}
        }
\altaffiltext{1}{McDonald Observatory and Dept.\ of Astronomy,
                 Univ.\ of Texas, Austin}
\altaffiltext{2}{Pulkovo Astronomical Observatory,
                 St.\,Petersburg, Russia}
\altaffiltext{3}{Sobolev Astronomical Institute,
                 St.\,Petersburg State University, Russia}
\altaffiltext{4}{Department of Astronomy, University of Michigan}
\altaffiltext{5}{School of Physics, UNSW, Sydney, Australia}
\altaffiltext{6}{Australian Centre for Astrobiology, UNSW, Sydney, Australia}

\begin{abstract}
Dynamical information along with survey data on metallicity and in some cases age have been used recently by some authors to search for candidates of stars that were born in the cluster where the Sun formed. We have acquired high resolution, high signal-to-noise ratio spectra for 30 of these objects to determine, using detailed elemental abundance analysis, if they could be true solar siblings. Only two of the candidates are found to have solar chemical composition. Updated modeling of the stars' past orbits in a realistic Galactic potential reveals that one of them, HD\,162826, satisfies both chemical and dynamical conditions for being a sibling of the Sun. Measurements of rare-element abundances for this star further confirm its solar composition, with the only possible exception of Sm. Analysis of long-term high-precision radial velocity data rules out the presence of hot Jupiters and confirms that this star is not in a binary system. We find that chemical tagging does not necessarily benefit from studying as many elements as possible, but instead from identifying and carefully measuring the abundances of those elements which show large star-to-star scatter at a given metallicity. Future searches employing data products from ongoing massive astrometric and spectroscopic surveys can be optimized by acknowledging this fact.
\end{abstract}

\keywords{stars: abundances ---
          stars: kinematics and dynamics ---
          stars: fundamental parameters ---
          stars: general ---
          stars: individual (HD\,162826)
}

\section{INTRODUCTION}

Infrared surveys and observations of young stars made over the past two decades suggest that most stars (80--90\,\%) are born in rich clusters of more than 100 members inside giant molecular clouds \citep{lada03, porras03,evans09}. Although it has been pointed out that, depending on the definition of cluster, the fraction of stars born in them could be as low as 45\,\% \citep{bressert10}, there is convincing evidence that the Sun was born in a moderately large stellar system.

Daughter products of short-lived ($<10$\,Myr) radioactive species have been found in meteorites, suggesting that the radioactive isotopes themselves were present in the early solar system \cite[e.g.,][]{goswami00,tachibana03}. A nearby supernova explosion could have injected these isotopes into the solar nebula \cite[e.g.,][]{looney06}, an event that has a high probability in a dense stellar environment \cite[e.g.,][]{williams07}. Additional evidence that the Sun was born in such an environment is provided by the dynamics of outer solar system objects like Sedna \citep{brown04}, which has large eccentricity ($e\sim0.8$) and perihelion ($\sim75$\,AU). Numerical simulations show that these extreme orbital properties can arise from close encounters with other stars \cite[e.g.,][]{morbidelli04}.

In his review of ``The Birth Environment of the Solar System,'' \cite{adams10} concludes that the Sun was born in a cluster of $10^3-10^4$ stars. The probability of close stellar encounters and nearby supernova pollution is high in a bound cluster with more than $\sim10^3$ members. On the other hand, the upper limit of $\sim10^4$ is set by the conditions that the solar system planets' orbits must be stable and that the early UV radiation fields were not strong enough to evaporate the solar nebula. \cite{pfalzner13} argues that the Sun most likely formed in an environment resembling an OB association (as opposed to a starburst cluster), where star densities are not sufficient for the destruction of protoplanetary disks due to stellar encounters.

The mean lifetime of open clusters in the Galactic disk is estimated to be about 100\,Myr \citep{janes88}. Considering the solar system age of 4.57\,Gyr \citep{bouvier10}, it is not surprising that the solar cluster is now fully dissipated and its members scattered throughout the Galaxy. The hypothesis that the Sun was born in the solar-age, solar-metallicity open cluster M67 \citep{johnson55,demarque92,randich06}, which hosts some of the most Sun-like stars known \citep{pasquini08,onehag11,onehag14}, has been refuted by \cite{pichardo12} using dynamical arguments. These authors even conclude that the Sun and M67 could not have been born in the same giant molecular cloud.

Stars that were born together with the Sun are referred to as ``solar siblings.'' They should not to be confused with ``solar twins,'' which are stars with high resolution, high signal-to-noise ratio spectra nearly indistinguishable from the solar spectrum, regardless of their origin \citep{porto97,melendez07:twins,ramirez11,monroe13}. By definition, the siblings of the Sun must have solar age and solar chemical composition because they formed essentially at the same time from the same gas cloud. They do not need to be ``Sun-like'' with respect to their fundamental parameters such as effective temperature, mass, luminosity, or surface gravity.

In principle, the siblings of the Sun could be found by measuring accurately and with high precision the ages and detailed chemical compositions of large samples of stars, an approach that is obviously impractical. Fortunately, analytical models of the Galactic potential can be used to predict their present-day dynamical properties. Alternatively, the same models can be employed to determine in retrospect if a given star could have possibly originated within the solar cluster. Thus, by applying dynamical constraints, manageable samples of solar sibling candidates can be constructed to be later examined carefully using more expensive methods such as high-resolution spectroscopy.

Thus, the Galactic mass distribution model by \cite{miyamoto75}, as given in \cite{paczynski90}, was used by \cite{portegies09} to reverse the orbit of the Sun and calculate its birth-place in the Galaxy. The orbits of simulated solar clusters with 2048 members were then computed to determine the fraction of solar siblings to be found today in the solar neighborhood. \citeauthor{portegies09} concludes that 10--60 of them should be found within 100\,pc from the Sun. This somewhat optimistic view was challenged by \cite{mishurov11}, who employed the Galactic potential by \cite{allen91} and perturbed it with both quasi-stationary \citep{lin69} and transient \citep{sellwood02} spiral density waves to show that the orbits of solar siblings strongly deviate from being nearly circular. \citeauthor{mishurov11} find that only a few solar siblings may be found within 100\,pc from the Sun if the solar cluster had $\sim10^3$ members. The situation might be worse if scattering by molecular clouds or close stellar encounters were to be taken into consideration, but ``less hopeless'' if the solar cluster contained $\sim10^4$ stars instead.

Following the modeling by \cite{portegies09}, \cite{brown10} pointed out that one could use the regions of phase-space occupied by the simulated stars to narrow-down the search for the siblings of the Sun. Their calculations suggest that these stars should have proper motions lower than 6.5\,mas\,yr$^{-1}$ and parallaxes larger than 10 mas. Then, by employing data from the {\it Hipparcos} catalog \citep{perryman97}, in its revised version \citep{vanleeuwen07}, they were able to find the 87 stars satisfying these conditions, after imposing an additional constraint of 10\,\% for the precision of the parallaxes. Their list of candidates was further narrowed down by excluding stars with $(B-V)<0.5$, which are bluer than the solar-age isochrone turn-off. Finally, they employed the stellar ages given in the Geneva-Copenhagen Survey \cite[GCS;][]{nordstrom04,holmberg07, holmberg09} to find the 6 stars with solar age within the errors. Based on the radial velocities and metallicities given in the GCS, \citeauthor{brown10} conclude that only one star (HD\,28676) may be a true solar sibling. Five other candidates with $(B-V)>1.0$ were not found in the GCS.

A procedure similar to the one described above was adopted by \cite{batista12}, but employing in addition to the GCS other metallicity data sets (namely \citealt{cayrel01}, \citealt{feltzing01}, and \citealt{valenti05}). By selecting stars with $-0.1<\feh<+0.1$ and using the same proper motion and parallax constraints described above, they created a list of 21 candidates for siblings of the Sun. Then, they used the previously published stellar atmospheric parameters and the PARAM code by \cite{dasilva06}\footnote{\url{http://stev.oapd.inaf.it/cgi-bin/param}} to estimate isochrone ages for those stars and thus find the nine objects with solar age within the errors. Further examination of the available data led them to conclude that the stars HD\,28676 (already suggested by \citealt{brown10}), HD\,83423, and HD\,175740 (the first giant proposed as candidate) could all well be true solar siblings.

The problem of searching for solar siblings in the solar neighborhood using existing survey data was tackled also by \cite{bobylev11}. They suggested a different method for finding them. First, targets were searched using the stars' Galactic space velocities $U,V,W$ (their data set is described in \citealt{bobylev10}) by excluding those whose total speeds relative to the solar one are significantly different ($\sqrt{U^2+V^2+W^2}>8$\,\kms). Then, the orbits of the remaining objects (162 FGK stars with parallax errors lower than 15\,\%) were simulated backwards in time using the \cite{allen91} Galactic potential perturbed by spiral density waves (as suggested by \citealt{mishurov11}) to determine parameters of encounter with the solar orbit such as relative distance and velocity difference over a period of time comparable to the Sun's age. Finally, these parameters were examined to calculate the probability that any of these objects was born together with the Sun. Their calculations rule out HD\,28676 and HD\,192324, both solar sibling candidates according to \cite{brown10}. On the other hand, \citeauthor{bobylev11} list as their best candidates the stars HD\,83423 and HD\,162826. Note that the former is also a good candidate according to \cite{batista12}. Furthermore, the calculated dynamic properties of this star seem to be very robust under different model parameters.

\begin{deluxetable*}{lccrrrrcc}
\centering
\tablewidth{0pt}
\tablecaption{Solar Sibling Candidates}
\tablehead{\colhead{HD} & \colhead{$V$} & \colhead{SpT\tablenotemark{a}} & $d$\tablenotemark{b} & $\sigma(d)$\tablenotemark{b} & \colhead{Observed\tablenotemark{c}} & \colhead{$RV$} & \colhead{$\sigma(RV)$} & \colhead{Reference\tablenotemark{d}} \\ & (mag) & & (pc) & (pc) & & (km\,s$^{-1}$) & (km\,s$^{-1}$) & }
\startdata
7735 & 7.91 & F5 & 85.7 & 8.8 & McD -- Dec.\,2012 & 28.6 & 0.2 & Ba12 \\
26690 & 5.29 & F2V & 36.3 & 0.8 & McD -- Dec.\,2012 & 16.0 & 0.1 & Ba12 \\
28676 & 7.09 & F5 & 38.7 & 0.9 & McD -- Dec.\,2012 & 6.4 & 0.1 & Br10+Ba12 \\
35317nw & 6.14 & F7V & 55.7 & 2.4 & McD -- Dec.\,2012 & 15.0 & 0.1 & Ba12 \\
44821 & 7.37 & K0V & 29.3 & 0.5 & McD -- Dec.\,2012 & 14.6 & 0.1 & Br10+Ba12 \\
46100 & 9.38 & G8V & 55.5 & 2.6 & LCO -- Apr.\,2013 & 21.0 & 0.3 & Ba12 \\
46301 & 7.28 & F5V & 107.6 & 6.6 & McD -- Dec.\,2012 & -4.7 & 0.1 & Ba12 \\
52242 & 7.41 & F2V & 68.2 & 2.7 & McD -- Dec.\,2012 & 31.3 & 0.9 & Ba12 \\
83423 & 8.78 & F8V & 72.1 & 4.9 & McD -- Dec.\,2012 & -4.1 & 0.1 & Bo11+Ba12 \\
91320 & 8.43 & G1V & 90.5 & 6.9 & McD -- Dec.\,2012 & 16.6 & 0.2 & Br10 \\
95915 & 7.25 & F6V & 66.6 & 2.1 & LCO -- Apr.\,2013 & 19.2 & 0.3 & Ba12 \\
100382 & 5.89 & K1III & 94.0 & 3.0 & LCO -- Apr.\,2013 & -8.5 & 0.3 & Br10 \\
101197 & 8.74 & G5 & 83.0 & 6.8 & McD -- Dec.\,2012 & 11.9 & 0.4 & Ba12 \\
102928 & 5.63 & K0III & 91.4 & 4.2 & McD -- Dec.\,2012 & 14.4 & 0.1 & Br10 \\
105000 & 7.91 & F2 & 71.1 & 3.0 & McD -- Dec.\,2012 & -13.8 & 0.1 & Ba12 \\
105678 & 6.34 & F6IV & 74.0 & 1.7 & McD -- Mar.\,2013 & -19.5 & 0.3 & Ba12 \\
147443 & 8.74 & G5 & 92.0 & 8.4 & McD -- Mar.\,2013 & 8.1 & 0.2 & Br10 \\
148317 & 6.70 & G0III & 79.6 & 3.5 & McD -- Mar.\,2013 & -37.9 & 0.1 & Ba12 \\
154747 & 8.73 & G8IV & 97.8 & 8.9 & LCO -- Apr.\,2013 & -14.9 & 0.3 & Ba12 \\
162826 & 6.56 & F8V & 33.6 & 0.4 & McD -- Dec.\,2012 & 1.7 & 0.1 & Bo11 \\
168442 & 9.66 & K7V & 19.6 & 0.6 & McD -- Dec.\,2012 & -14.3 & 0.2 & Br10 \\
168769 & 9.37 & K0V & 50.2 & 3.7 & LCO -- Apr.\,2013 & 26.3 & 0.2 & Br10 \\
175740 & 5.44 & G8III & 82.0 & 1.7 & McD -- Dec.\,2012 & -9.8 & 0.1 & Br10+Ba12 \\
183140 & 9.30 & G8V & 71.8 & 6.6 & LCO -- Apr.\,2013 & -30.0 & 0.1 & Ba12 \\
192324 & 7.66 & F8 & 67.1 & 4.8 & LCO -- Apr.\,2013 & -3.8 & 0.2 & Br10 \\
196676 & 6.45 & K0 & 74.4 & 2.8 & McD -- Dec.\,2012 & -0.5 & 0.2 & Br10 \\
199881 & 7.49 & F5 & 72.2 & 3.6 & McD -- Dec.\,2012 & -16.7 & 0.1 & Ba12 \\
199951 & 4.67 & G6III & 70.2 & 1.3 & McD -- Dec.\,2012 & 18.2 & 0.2 & Ba12 \\
207164 & 7.54 & F2 & 76.1 & 3.8 & McD -- Dec.\,2012 & -6.7 & 0.1 & Ba12 \\
219828 & 8.05 & G0IV & 72.3 & 3.9 & McD -- Dec.\,2012 & -24.3 & 0.1 & Ba12 

\enddata
\tablenotetext{a}{Spectral type from SIMBAD.}
\tablenotetext{b}{Distance derived from \textit{Hipparcos} parallax.}
\tablenotetext{c}{McD: Tull spectrograph on the 2.7\,m Harlan J.~Smith Telescope at McDonald Observatory. LCO: Mike spectrograph on the 6.5\,m Magellan/Clay Telescope at Las Campanas Observatory.}
\tablenotetext{d}{Br10: \cite{brown10}, Bo11: \cite{bobylev11}, Ba12: \cite{batista12}.}
\label{t:sample}
\end{deluxetable*}

In this work, we perform chemical analysis using newly acquired high resolution, high signal-to-noise ratio spectra to establish if any of the stars discussed above can be considered a true solar sibling. Finding these objects could shed light onto our origins. For example, knowledge of the spatial distribution of solar siblings can help determining the Sun's birthplace and understand the importance of radial mixing in shaping the properties of disk galaxies \citep{bland-hawthorn10}. It will also enable us to better constrain the physical characteristics of the environment in which our star was born \citep{adams10}. Furthermore, theoretical calculations show that large impacts of rocks into planets produce fragments where primitive life can survive as they travel to other planets or even other nearby planetary systems, a mechanism known as ``lithopanspermia'' \citep{adams05,valtonen09,belbruno12}. For example, \cite{worth13} estimate that about 5\,\% of impact remains from Earth have escaped the solar system. A similar value is obtained for the case of Mars. \cite{valtonen12} point out that the possibility of contamination of solar siblings by ``spores of life'' from our planet should make the siblings of the Sun prime targets in the search for extraterrestrial life.  

As we enter the era of massive, far-reaching astrometric/photometric surveys led by Gaia \citep{lindegren12} as well as comparably large high-resolution spectroscopic surveys such as Gaia-ESO \citep{gilmore12} and GALAH \citep{zucker12}, a thorough investigation of the available data, albeit relatively small in the context of solar sibling research, must be made to guide the more statistically significant search strategies of the near future.

\section{SAMPLE AND SPECTROSCOPIC ANALYSIS}

\subsection{The Solar Sibling Candidates}

Our target list consists of all interesting objects discussed in the previous exploratory searches by \citet[hereafter Br10]{brown10}, \citet[hereafter Bo11]{bobylev11}, and \citet[hereafter Ba12]{batista12}.\footnote{Very recently, \cite{batista14} employed the detailed elemental abundance analysis of one of the largest exoplanet host star samples to find another solar sibling candidate: HD\,186302. We did not include this star in our work.} In addition to the six candidates listed in Table~1 of Br10, i.e., those they found in the GCS catalog, we also observed the five stars with $(B-V)>1.0$ that satisfied all other criteria for solar sibling candidate in that work. The two stars discussed in detail in Bo11 were also included. Finally, all 21 stars in the extended list by Ba12 (see their Table~A) were observed. Note that Ba12's list includes one object from Bo11 and three from Br11. Thus, a total of 30 targets were observed for this work (Table~\ref{t:sample}).

\subsection{Spectroscopic Observations}

We used the Tull coud\'e spectrograph on the 2.7\,m Harlan J.\ Smith Telescope at McDonald Observatory \citep{tull95} to observe most of our targets (23). All but three of them were observed in December 2012; the others were observed in March 2013. The rest of our targets (7) have too low declinations to be observed from McDonald Observatory. Instead, they were observed using the MIKE spectrograph on the 6.5\,m Telescope at Las Campanas Observatory \citep{bernstein03} in April 2013. Slit sizes were chosen so that the spectral resolution of the data is about 60\,000 in the visible part of the spectrum. We targeted a signal-to-noise ratio ($S/N$) per pixel of at least 200 at 6\,000\,\AA. Only one of our targets (HD\,46100) has a significantly lower $S/N$ spectrum.

The Tull/McDonald spectra were reduced in the standard manner using the echelle package in IRAF\footnote{IRAF is distributed by the National Optical Astronomy Observatory, which is operated by the Association of Universities for Research in Astronomy (AURA) under cooperative agreement with the National Science Foundation.} while the MIKE/Magellan data were reduced using the CarnegiePython pipeline.\footnote{\url{http://code.obs.carnegiescience.edu/mike}} After correcting for the Earth's motion using the rvcorrect task in IRAF, the radial velocities ($RV$s) of our target stars were measured by cross-correlation with the spectra of stars in our sample with known stable radial velocities. For the McDonald data we used as radial velocity standards the stars HD\,196676 ($RV=-0.5$\,\kms) and HD\,219828 ($RV=-24.1$\,\kms). Their radial velocities are from \cite{chubak12}. Three of the stars observed from Las Campanas have radial velocities in the GCS catalog (HD\,46100, $RV=21.0\pm0.2$\,\kms; HD\,183140, $RV=-29.3\pm0.1$\,\kms, and HD\,192324, $RV=-4.4\pm0.2$). In all cases, regardless of which star(s) is(are) used as standards, the resulting $RV$ values are robust. They are given in Table~\ref{t:sample}. Formal, internal errors on our $RV$ values are typically around 0.2\,\kms. These errors are due to the order-to-order cross-correlation velocity scatter and the uncertainties in the velocities of the standard stars. Systematic uncertainties will increase these errors by at least 0.5\,\kms.

\subsection{Moderately Fast Rotators and Double-lined Spectroscopic Binaries}

\begin{figure}
\centering
\includegraphics[bb=88 365 394 745,width=8cm]{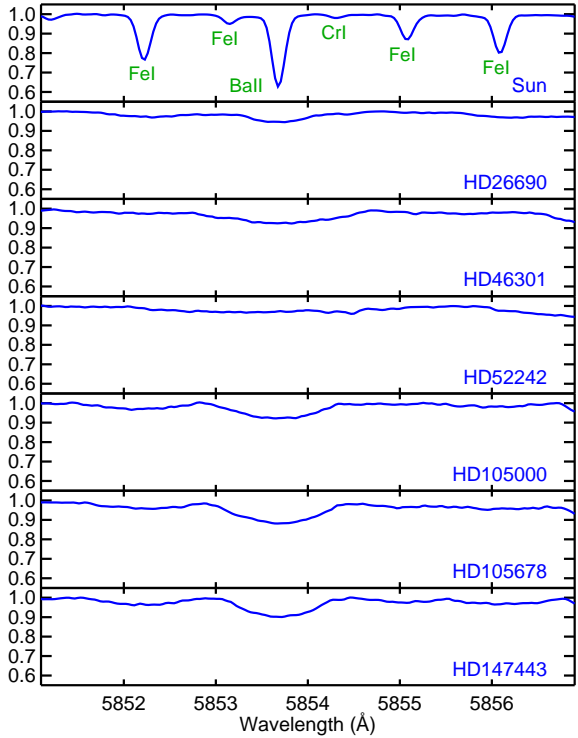}
\caption{Spectra of high $V\sin i$ stars. Our solar (Vesta) spectrum is shown in the top panel for comparison.}
\label{f:high_vsini}
\end{figure}

The methods that we employ to derive atmospheric parameters and to measure elemental abundances (see Sections~\ref{s:pars} and \ref{s:abundances}) are not suited for stars with high projected rotational velocity ($V\sin i$) or double-lined spectroscopic binaries (SB2s). The spectral lines in the former are either severely blended or they cannot be identified due to the extreme rotational broadening. SB2s, on the other hand, require special treatment because even in cases where both sets of spectral lines can be identified and measured independently, the blended continuum flux must be first estimated from previous knowledge of the two stars, making the problem somewhat degenerate and the results not as accurate as those that can be achieved for single spectra stars. More importantly, in most cases these stars have been selected because their photometry suggests a solar metallicity. Photometric calibrations of metallicity should only be applied to single stars.

\begin{figure}
\centering
\includegraphics[bb=88 365 394 745,width=8cm]{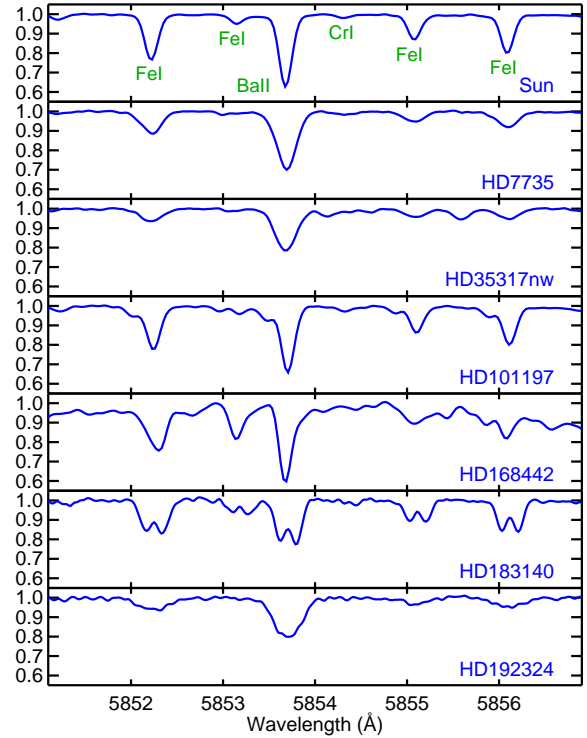}
\caption{Spectra of double-lined spectroscopic binary (SB2) stars. Our solar (Vesta) spectrum is shown in the top panel for comparison.}
\label{f:sb2}
\end{figure}

Figures~\ref{f:high_vsini} and \ref{f:sb2} show the small spectral region of our spectroscopic data containing the 5853.7\,\AA\ Ba~\textsc{ii} line for our target stars with very high $V\sin i$ and targets which are SB2s, respectively. For reference, our solar (Vesta) spectrum is shown in the top panel of these figures. The SB2 nature of HD\,101197, and specially HD\,183140 is obvious. The spectra of HD\,7735 and HD\,35317 show excess absorption on the blue wings of all spectral lines. The latter in fact has a very high $V\sin i$ companion, which was not analyzed in this work (hence the ``nw'' flag, which stands for North-West component). HD\,168442 has a spectrum that is irreproducible with single star models. It appears to be a blend of a Sun-like star and an M-type dwarf. HD\,192324 has a very nearby companion which appears to have contaminated our spectrum. Even though this star can be analyzed and stellar parameters were determined as described in Section~\ref{s:pars}, a very clear trend of iron abundance with wavelength persists, suggesting that the line strengths have been affected by the contribution to the continuum of the nearby, and probably very cool companion.

All moderately fast rotators and SB2s were excluded from further analysis. Therefore, hereafter our sample is reduced to 18 stars.

\subsection{Stellar Parameter Determination} \label{s:pars}

We employed the standard technique of excitation/ionization balance of iron lines to determine the stars' atmospheric parameters $\teff$ (effective temperature), $\logg$ (surface gravity), $\feh$ (iron abundance), and $\vt$ (microturbulence). The details of this method have been described multiple times, for example in \cite{ramirez09,ramirez11,ramirez13:thin-thick}. Basically, the equivalent widths of a large number of non-blended, unsaturated \fei\ and \feii\ lines are measured using Gaussian line profile fits. In our case, this was done using IRAF's splot task. Then, a standard curve-of-growth approach is employed to determine the iron abundance from each line for a given set of guess stellar parameters.

We used the spectrum synthesis code MOOG \citep{sneden73}\footnote{\url{http://www.as.utexas.edu/~chris/moog.html}} and MARCS model atmospheres with standard chemical composition \citep{gustafsson08}\footnote{Available online at \url{http://marcs.astro.uu.se}} for our iron abundance calculations. The initial guess stellar parameters were iteratively modified until the correlations between iron abundance and excitation potential and reduced equivalent width of the line disappear while simultaneously enforcing agreement between the mean iron abundances inferred from \fei\ and \feii\ lines separately. For these calculations, we employed relative iron abundances, i.e., the iron abundances were measured differentially with respect to the Sun, on a line-by-line basis, and using the reflected sunlight spectrum of the asteroid Vesta, taken from McDonald Observatory in December 2012.

Formal errors for our spectroscopically derived stellar parameters were calculated as described in Appendix~B of \cite{bensby14} and in Section~3.2 of \cite{epstein10}. These errors are small given the high-quality of our data (hence highly-precise $EW$ values) and the strict differential nature of our work which minimizes the impact of uncertainties in the atomic data. Indeed, the average errors in $\teff$, $\logg$, and $\feh$ are 30\,K, 0.07\,dex, and 0.02\,dex, respectively. However, we note that these values are not representative of the real errors in these parameters, which are dominated by systematic uncertainties, as will be discussed in Section~\ref{s:systematics}.

\begin{deluxetable}{lcll}
\centering
\tablecaption{Iron Line List}
\tablehead{\colhead{Wavelength} & \colhead{Species\tablenotemark{a}} & \colhead{EP} & \colhead{$\log gf$} \\ \colhead{(\AA)} & \colhead{} & \colhead{(eV)} & \colhead{}}
\startdata
4779.44 & 26.0 & 3.42 & -2.16 \\
4788.76 & 26.0 & 3.24 & -1.73 \\
4799.41 & 26.0 & 3.64 & -2.13 \\
4808.15 & 26.0 & 3.25 & -2.69 \\
4961.91 & 26.0 & 3.63 & -2.19 \\
5044.211 & 26.0 & 2.851 & -2.058 \\
5054.64 & 26.0 & 3.64 & -1.98 \\
5187.91 & 26.0 & 4.14 & -1.26 \\
5197.94 & 26.0 & 4.3 & -1.54 \\
5198.71 & 26.0 & 2.22 & -2.14 \\

\vdots & \vdots & \vdots & \vdots 
\enddata
\tablenotetext{a}{The number to the left of the decimal point indicates the atomic number. The number to the right of the decimal point indicates the ionization state, where ``0'' is neutral and ``1'' is singly ionized.}
\label{t:linelist_iron}
\end{deluxetable}

\begin{deluxetable*}{lccrccr}
\centering
\tablecaption{Stellar Parameters}
\tablehead{\colhead{HD} & \colhead{$\teff$} & \colhead{$\logg$} & \colhead{$\feh$} & \colhead{$\teff'$} & \colhead{$\logg'$} & \colhead{$\feh'$} \\ \colhead{} & \colhead{(K)} & \colhead{[cgs]} & \colhead{} & \colhead{(K)} & \colhead{[cgs]} & \colhead{}}
\startdata
28676 & $5942\pm10$ & $4.32\pm0.03$ & $0.09\pm0.01$ & $5845\pm27$ & $4.22\pm0.03$ & $0.04\pm0.03$ \\
44821 & $5727\pm11$ & $4.52\pm0.06$ & $0.04\pm0.03$ & $5743\pm23$ & $4.50\pm0.02$ & $0.04\pm0.03$ \\
46100 & $5543\pm41$ & $4.58\pm0.02$ & $0.04\pm0.01$ & $5488\pm32$ & $4.55\pm0.02$ & $0.02\pm0.08$ \\
83423 & $6096\pm16$ & $4.48\pm0.04$ & $0.00\pm0.01$ & $6090\pm36$ & $4.43\pm0.04$ & $-0.01\pm0.04$ \\
91320 & $6146\pm18$ & $4.23\pm0.04$ & $0.11\pm0.01$ & $5975\pm30$ & $4.11\pm0.05$ & $0.02\pm0.05$ \\
95915 & $6414\pm41$ & $4.22\pm0.08$ & $-0.01\pm0.02$ & $6280\pm42$ & $4.02\pm0.03$ & $-0.08\pm0.05$ \\
100382 & $4751\pm63$ & $2.69\pm0.16$ & $0.08\pm0.08$ & $4603\pm53$ & $2.56\pm0.07$ & $0.12\pm0.11$ \\
102928 & $4796\pm32$ & $2.84\pm0.12$ & $-0.08\pm0.03$ & $4615\pm102$ & $2.48\pm0.13$ & $-0.12\pm0.08$ \\
148317 & $6041\pm17$ & $3.87\pm0.04$ & $0.21\pm0.01$ & $5859\pm23$ & $3.63\pm0.03$ & $0.11\pm0.05$ \\
154747 & $5322\pm17$ & $3.87\pm0.05$ & $0.00\pm0.02$ & $5268\pm22$ & $3.86\pm0.05$ & $-0.01\pm0.04$ \\
162826 & $6210\pm13$ & $4.41\pm0.03$ & $0.03\pm0.01$ & $6101\pm32$ & $4.25\pm0.02$ & $-0.04\pm0.03$ \\
168769 & $5355\pm16$ & $4.42\pm0.04$ & $0.03\pm0.01$ & $5218\pm33$ & $4.54\pm0.05$ & $0.14\pm0.10$ \\
175740 & $4890\pm22$ & $2.91\pm0.09$ & $0.09\pm0.03$ & $4784\pm61$ & $2.73\pm0.06$ & $0.09\pm0.06$ \\
196676 & $4841\pm47$ & $3.10\pm0.10$ & $0.04\pm0.04$ & $4808\pm80$ & $3.03\pm0.10$ & $0.01\pm0.07$ \\
199881 & $6691\pm62$ & $4.35\pm0.14$ & $0.16\pm0.04$ & $6546\pm41$ & $4.20\pm0.04$ & $0.09\pm0.08$ \\
199951 & $5218\pm36$ & $3.09\pm0.08$ & $-0.04\pm0.03$ & $5129\pm55$ & $2.86\pm0.03$ & $-0.10\pm0.07$ \\
207164 & $6886\pm61$ & $4.47\pm0.12$ & $0.22\pm0.03$ & $6742\pm32$ & $4.24\pm0.04$ & $0.14\pm0.06$ \\
219828 & $5886\pm13$ & $4.18\pm0.03$ & $0.19\pm0.01$ & $5815\pm22$ & $4.12\pm0.05$ & $0.15\pm0.03$ 

\enddata
\label{t:pars}
\end{deluxetable*}

Our iron line list and adopted atomic data are given in Table~\ref{t:linelist_iron}. The stellar parameters derived as described above (hereafter referred to as the ``spectroscopic'' parameters $\teff$, $\logg$, $\feh$) are listed in Table~\ref{t:pars}. The other set of parameters listed in Table~\ref{t:pars}, $\teff'$, $\logg'$, $\feh'$ will be described in Section~\ref{s:systematics}.

\subsection{Reliability of Photometric Metallicities}

\begin{figure}
\centering
\includegraphics[bb=85 235 535 565,width=8.2cm]{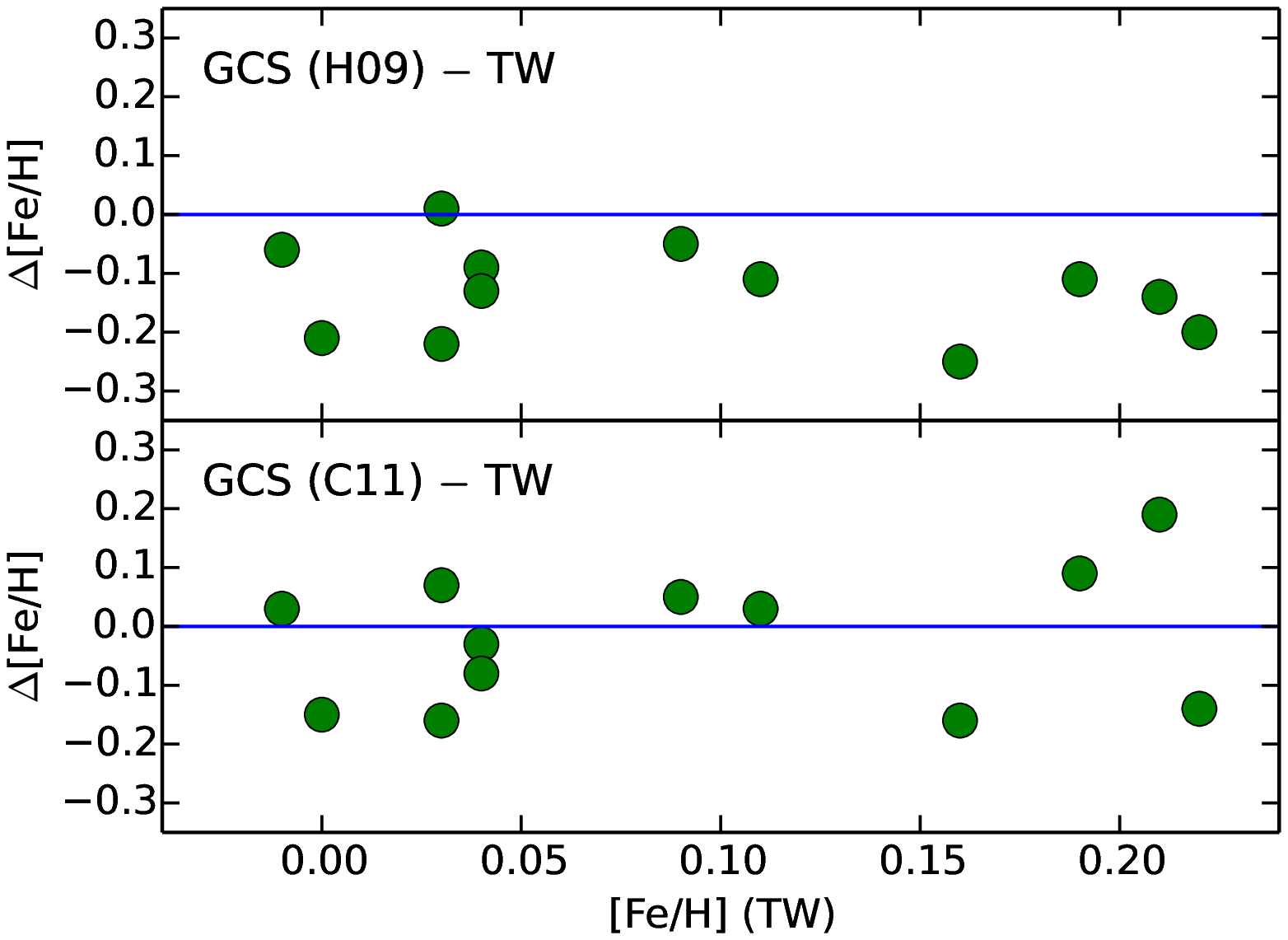}
\caption{Comparison of iron abundances. Top panel: $\feh$ values from the GCS \cite[][H09]{holmberg09} minus those derived in this work (TW) as a function of $\feh$\,(TW). Bottom panel: as in the top panel for the improved GCS values by \cite[][C11]{casagrande11}.}
\label{f:gcs}
\end{figure}

A comparison of our spectroscopically-derived $\feh$ values with those inferred from photometric calibrations, as given in the GCS catalog, is presented in Figure~\ref{f:gcs}. This comparison is relevant because solar sibling searches can benefit from a reasonable $\feh$ constraint, and large catalogs like the GCS have been found convenient for that purpose. The top panel in Figure~\ref{f:gcs} shows that the GCS $\feh$ values, as given in \cite{holmberg09}, are systematically low by about 0.1\,dex relative to ours (the mean difference is $-0.13\pm0.07$; the error bar here corresponds to the 1-$\sigma$ star-to-star scatter). This, combined with the fact that most previous exploratory searches of solar siblings have employed the original GCS catalog, is the reason why our sample centers around $\feh\sim+0.1$ and not $\feh=0$.

The bottom panel of Figure~\ref{f:gcs} shows that the improved GCS metallicities given in \cite{casagrande11} are consistent with our spectroscopic solutions. \cite{casagrande11} have discussed at length the systematic offset required for the original GCS $\feh$ values, which essentially stems from a better and more consistent set of $\teff$ values in the underlying photometric metallicity calibration. The average difference in $\feh$ values between those given by \cite{casagrande11} and ours is $-0.02\pm0.11$.

The discussion above regarding Figure~\ref{f:gcs} shows that the \cite{casagrande11} GCS metallicities should be the preferred set for the purpose of constraining a stellar sample based on the stars' $\feh$ values. Moreover, although the 1-$\sigma$ errors of the photometric metallicities are quoted typically as 0.1\,dex, one should keep in mind that this number corresponds to a sample and not to individual stars. By definition, at least 30\,\% of stars with real $|\feh|<0.1$ have photometric $\feh$ values outside of that ``solar'' range. Thus, a metallicity constraint of 0.1\,dex already excludes a good number of potentially good candidates. Given that in the case of a search for nearby siblings of the Sun it is crucial not to discard a single potentially interesting candidate, perhaps the safer choice should be $|\feh|<0.2$, which would exclude only about 5\,\% of real solar-metallicity stars. Indeed, none of the 5 key solar sibling candidates that will be discussed in Section~\ref{s:key} would have survived a $|\feh|<-0.1$ cut had we used the \cite{casagrande11} GCE metallicities.

\subsection{Elemental Abundance Determination} \label{s:abundances}

We employed equivalent width measurements and standard curve-of-growth analysis to derive the abundances of 14 elements other than iron: O, Na, Al, Si, Ca, Sc, Ti, V, Cr, Mn, Co, Ni, Y, and Ba. As in the case of iron, equivalent widths were measured using IRAF's splot task while the curve-of-growth analysis was made using MOOG and MARCS model atmospheres. Oxygen abundances were inferred from the 777\,nm O\,\textsc{i} triplet lines and corrected for departures from local thermodynamical equilibrium (LTE) using the grid of non-LTE corrections by \cite{ramirez07}.\footnote{An online tool to calculate these non-LTE corrections is available at \url{http://www.as.utexas.edu/~ivan}} Hyperfine structure was taken into account for lines due to V, Mn, Co, Y, and Ba using the wavelengths and relative $\log gf$ values from the Kurucz atomic line database.\footnote{\url{http://kurucz.harvard.edu/linelists.html}} Our adopted linelist for elements other than iron is given in Table~\ref{t:linelist_other}. Our derived abundances are listed in Tables~\ref{t:abundances_m} and \ref{t:abundances_notm}. Errors listed in these tables correspond to the 1-$\sigma$ line-to-line scatter and do not include systematic uncertainties. The latter will be discussed in Section~\ref{s:systematics}.

\begin{deluxetable}{lrlr}
\centering
\tablecaption{Line List For Elements Other Than Iron}
\tablehead{\colhead{Wavelength} & \colhead{Species\tablenotemark{a}} & \colhead{EP} & \colhead{$\log gf$} \\ \colhead{(\AA)} & \colhead{} & \colhead{(eV)} & \colhead{}}
\startdata
7771.9438 & 8.0 & 9.146 & 0.352 \\
7774.1611 & 8.0 & 9.146 & 0.223 \\
7775.3901 & 8.0 & 9.146 & 0.002 \\
5688.21 & 11.0 & 2.1 & -0.48 \\
6154.2251 & 11.0 & 2.102 & -1.547 \\
6160.7471 & 11.0 & 2.104 & -1.246 \\
5557.07 & 13.0 & 3.14 & -2.21 \\
6696.0181 & 13.0 & 3.143 & -1.481 \\
6698.667 & 13.0 & 3.143 & -1.782 \\
7835.309 & 13.0 & 4.022 & -0.689 \\

\vdots & \vdots & \vdots & \vdots 
\enddata
\tablenotetext{a}{The number to the left of the decimal point indicates the atomic number. The number to the right of the decimal point indicates the ionization state, where ``0'' is neutral and ``1'' is singly ionized.}
\label{t:linelist_other}
\end{deluxetable}

\begin{deluxetable*}{lrrrrrrrrr}
\centering
\tablewidth{0pt}
\tablecaption{Abundances of ``M'' Elements: O, Si, Ca, Sc, Ti, Cr, Mn, Co, and Ni}
\tablehead{\colhead{HD} & \colhead{[O/H]} & \colhead{[Si/H]} & \colhead{[Ca/H]} & \colhead{[Sc/H]} & \colhead{[Ti/H]} & \colhead{[Cr/H]} & \colhead{[Mn/H]} & \colhead{[Co/H]} & \colhead{[Ni/H]}}
\startdata
28676 & $0.05\pm0.03$ & $0.10\pm0.02$ & $0.09\pm0.03$ & $0.14\pm0.02$ & $0.10\pm0.01$ & $0.08\pm0.02$ & $0.10\pm0.01$ & $0.11\pm0.04$ & $0.10\pm0.03$ \\
44821 & $0.03\pm0.02$ & $0.01\pm0.02$ & $0.03\pm0.02$ & $-0.02\pm0.03$ & $0.02\pm0.01$ & $0.06\pm0.03$ & $0.03\pm0.02$ & $-0.03\pm0.05$ & $-0.01\pm0.03$ \\
46100 & $-0.12\pm0.03$ & $0.02\pm0.05$ & $0.02\pm0.03$ & $0.07\pm0.10$ & $0.00\pm0.03$ & $0.08\pm0.06$ & $0.00\pm0.03$ & $0.01\pm0.06$ & $0.01\pm0.05$ \\
83423 & $-0.01\pm0.03$ & $-0.02\pm0.03$ & $0.00\pm0.04$ & $-0.06\pm0.02$ & $0.00\pm0.02$ & $-0.04\pm0.03$ & $-0.08\pm0.02$ & $-0.01\pm0.05$ & $-0.09\pm0.03$ \\
91320 & $0.11\pm0.02$ & $0.12\pm0.03$ & $0.11\pm0.04$ & $0.11\pm0.04$ & $0.09\pm0.01$ & $0.08\pm0.04$ & $0.11\pm0.03$ & $0.16\pm0.02$ & $0.10\pm0.04$ \\
95915 & $-0.03\pm0.01$ & $0.01\pm0.05$ & $0.09\pm0.06$ & $-0.04\pm0.04$ & $0.03\pm0.06$ & $-0.01\pm0.02$ & $-0.10\pm0.05$ & $0.08\pm0.07$ & $-0.05\pm0.06$ \\
100382 & $0.08\pm0.08$ & $0.23\pm0.09$ & $0.06\pm0.06$ & $0.12\pm0.05$ & $0.08\pm0.05$ & $0.08\pm0.06$ & $0.07\pm0.07$ & $0.16\pm0.03$ & $0.13\pm0.07$ \\
102928 & $-0.08\pm0.03$ & $0.06\pm0.05$ & $-0.06\pm0.04$ & $-0.03\pm0.04$ & $-0.06\pm0.01$ & $-0.09\pm0.06$ & $-0.15\pm0.02$ & $-0.04\pm0.05$ & $-0.08\pm0.05$ \\
148317 & $0.09\pm0.02$ & $0.22\pm0.03$ & $0.21\pm0.03$ & $0.26\pm0.04$ & $0.24\pm0.01$ & $0.20\pm0.03$ & $0.23\pm0.03$ & $0.23\pm0.05$ & $0.22\pm0.04$ \\
154747 & $-0.01\pm0.02$ & $0.02\pm0.03$ & $0.00\pm0.03$ & $0.03\pm0.03$ & $0.05\pm0.01$ & $0.02\pm0.05$ & $-0.01\pm0.03$ & $0.00\pm0.04$ & $0.02\pm0.04$ \\
162826 & $0.01\pm0.02$ & $0.04\pm0.02$ & $0.02\pm0.03$ & $0.06\pm0.03$ & $0.04\pm0.01$ & $0.01\pm0.02$ & $0.01\pm0.03$ & $0.01\pm0.02$ & $0.01\pm0.03$ \\
168769 & $0.05\pm0.01$ & $0.03\pm0.03$ & $0.07\pm0.04$ & $-0.07\pm0.01$ & $0.03\pm0.02$ & $0.09\pm0.05$ & $0.04\pm0.05$ & $-0.04\pm0.04$ & $-0.01\pm0.03$ \\
175740 & $0.09\pm0.03$ & $0.24\pm0.05$ & $0.09\pm0.05$ & $0.12\pm0.03$ & $0.07\pm0.01$ & $0.08\pm0.05$ & $0.06\pm0.05$ & $0.10\pm0.06$ & $0.10\pm0.05$ \\
196676 & $0.09\pm0.04$ & $0.14\pm0.06$ & $0.05\pm0.04$ & $0.06\pm0.04$ & $0.02\pm0.01$ & $0.00\pm0.05$ & $0.01\pm0.02$ & $0.06\pm0.05$ & $0.07\pm0.06$ \\
199881 & $0.23\pm0.07$ & $0.16\pm0.03$ & $0.12\pm0.04$ & \nodata & $0.15\pm0.02$ & $0.06\pm0.03$ & $0.14\pm0.08$ & $0.20\pm0.01$ & $0.13\pm0.05$ \\
199951 & $-0.01\pm0.03$ & $0.02\pm0.05$ & $-0.03\pm0.04$ & $0.00\pm0.03$ & $0.01\pm0.03$ & $-0.11\pm0.01$ & $-0.14\pm0.03$ & $-0.02\pm0.05$ & $-0.09\pm0.06$ \\
207164 & $0.31\pm0.06$ & $0.28\pm0.04$ & $0.25\pm0.05$ & $0.35\pm0.12$ & $0.24\pm0.02$ & $0.18\pm0.02$ & $0.18\pm0.04$ & $0.39\pm0.01$ & $0.24\pm0.05$ \\
219828 & $0.14\pm0.02$ & $0.21\pm0.03$ & $0.18\pm0.02$ & $0.29\pm0.03$ & $0.25\pm0.01$ & $0.18\pm0.03$ & $0.20\pm0.01$ & $0.24\pm0.06$ & $0.22\pm0.03$ 

\enddata
\label{t:abundances_m}
\end{deluxetable*}

\begin{deluxetable*}{lrrrrr}
\centering
\tablecaption{Abundances of ``not M'' Elements: Na, Al, V, Y, and Ba}
\tablehead{\colhead{HD} & \colhead{[Na/H]} & \colhead{[Al/H]} & \colhead{[V/H]} & \colhead{[Y/H]} & \colhead{[Ba/H]}}
\startdata
28676 & $0.11\pm0.01$ & $0.08\pm0.02$ & $0.09\pm0.03$ & $0.09\pm0.06$ & $0.09\pm0.03$ \\
44821 & $-0.05\pm0.01$ & $-0.03\pm0.04$ & $0.02\pm0.04$ & $0.10\pm0.04$ & $0.14\pm0.01$ \\
46100 & $-0.05\pm0.04$ & $0.02\pm0.08$ & $0.17\pm0.07$ & $0.16\pm0.05$ & $0.15\pm0.04$ \\
83423 & $-0.12\pm0.04$ & $-0.21\pm0.05$ & $-0.03\pm0.08$ & $0.09\pm0.04$ & $0.26\pm0.03$ \\
91320 & $0.16\pm0.03$ & $0.07\pm0.04$ & $0.25\pm0.08$ & $0.10\pm0.03$ & $0.10\pm0.03$ \\
95915 & $0.04\pm0.06$ & $-0.08\pm0.01$ & $0.16\pm0.12$ & $0.05\pm0.03$ & $0.12\pm0.05$ \\
100382 & $0.25\pm0.07$ & $0.25\pm0.07$ & $0.20\pm0.03$ & $0.17\pm0.05$ & $0.04\pm0.04$ \\
102928 & $0.00\pm0.03$ & $0.02\pm0.01$ & $-0.02\pm0.04$ & $0.12\pm0.03$ & $0.07\pm0.05$ \\
148317 & $0.23\pm0.04$ & $0.18\pm0.03$ & $0.23\pm0.03$ & $0.28\pm0.02$ & $0.21\pm0.01$ \\
154747 & $0.00\pm0.04$ & $0.01\pm0.03$ & $0.06\pm0.04$ & $0.03\pm0.07$ & $-0.03\pm0.02$ \\
162826 & $0.02\pm0.03$ & $-0.04\pm0.05$ & $0.03\pm0.05$ & $0.04\pm0.04$ & $0.09\pm0.03$ \\
168769 & $-0.07\pm0.05$ & $0.01\pm0.03$ & $0.07\pm0.03$ & $0.14\pm0.10$ & $0.17\pm0.01$ \\
175740 & $0.24\pm0.08$ & $0.16\pm0.02$ & $0.12\pm0.05$ & $0.23\pm0.02$ & $0.14\pm0.02$ \\
196676 & $0.08\pm0.07$ & $0.15\pm0.01$ & $0.10\pm0.05$ & $0.24\pm0.07$ & $0.07\pm0.04$ \\
199881 & $0.36\pm0.01$ & \nodata & \nodata & $0.28\pm0.06$ & $0.21\pm0.11$ \\
199951 & $0.16\pm0.03$ & $-0.06\pm0.02$ & $0.04\pm0.04$ & $0.04\pm0.06$ & $0.31\pm0.05$ \\
207164 & $0.31\pm0.06$ & \nodata & \nodata & $0.39\pm0.03$ & $0.29\pm0.06$ \\
219828 & $0.22\pm0.01$ & $0.22\pm0.02$ & $0.20\pm0.03$ & $0.15\pm0.02$ & $0.16\pm0.03$ 

\enddata
\label{t:abundances_notm}
\end{deluxetable*}

In addition to Fe, lines due to neutral and singly-ionized Ti and Cr are available in the spectra of our stars. Thus, we derived Ti and Cr abundances using Ti\,\textsc{i} and Ti\,\textsc{ii} as well as Cr\,\textsc{i} and Cr\,\textsc{ii} lines separately in each case. The mean differences in Ti and Cr abundances inferred from the two types of lines are [Ti\,\textsc{i}/H]$-$[Ti\,\textsc{ii}/H]$=-0.03\pm0.06$ and [Cr\,\textsc{i}/H]$-$[Cr\,\textsc{ii}/H]$=+0.02\pm0.05$, i.e., consistent with zero within the observational uncertainties, but not exactly zero, as one would expect if true ionization balance had been achieved. The latter reflects our limitations in the modeling of stellar atmospheres and spectral line formation. Nevertheless, given the opposite signs of the Ti and Cr differences, we do not expect improved models to be significantly different from the ones employed in this work.

\section{LOOKING FOR THE SUN'S SIBLINGS}

\begin{figure}
\centering
\includegraphics[bb=115 235 502 580,width=8cm]{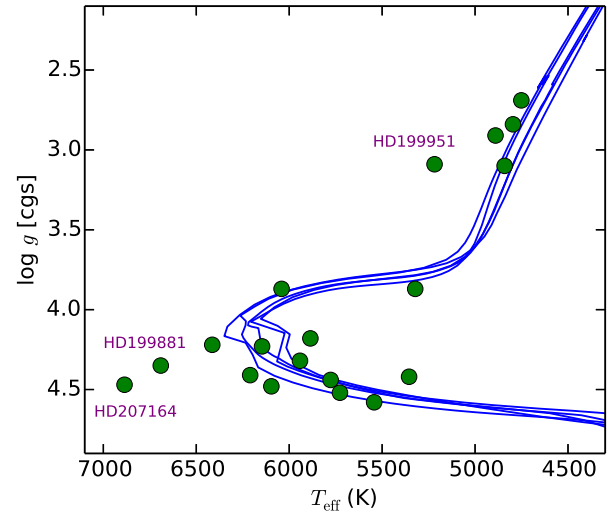}
\caption{Effective temperature versus surface gravity. Various theoretical isochrones of solar age and solar metallicity have been overplotted. The three stars whose parameters are not compatible with the solar-age, solar-metallicity isochrone are labeled.}
\label{f:hr}
\end{figure}

\subsection{The Solar-Age, Solar-Metallicity Isochrone}

Figure~\ref{f:hr} shows one version of the theoretical Hertzprung-Russell diagram using our derived stellar parameters effective temperature and surface gravity. It also shows several theoretical isochrones of solar age ($\sim4.5$\,Gyr) and solar metallicity. Although different isochrones have different definitions of solar metallicity, the differences are small and not important for our purposes. Isochrones computed by the following groups are shown in Figure~\ref{f:hr} ordered by their hottest isochrone point (coolest is last in this list): \cite{worthey94,bertelli94,dotter08,pietrinferni04,yi01}. There are important differences between them, but collectively they can help us discard a few stars based on a zeroth-order age estimate.

Determining ages of individual stars is a very difficult task, particularly for stars on the main-sequence, but a quick inspection of Figure~\ref{f:hr} clearly shows that three of our sample stars cannot have solar age within any reasonable uncertainties. HD\,199881 and HD\,207164 are too warm given the turn-off $\teff$ of the solar-age, solar-metallicity isochrone, which is at most $\sim6300$\,K. HD\,199951, on the other hand, appears to be a giant star of younger age than solar. All of our other targets have stellar parameters reasonably consistent with the solar-age, solar-metallicity isochrone.

\subsection{Chemical Tagging} \label{s:chemtag}

Although $\feh$ values are an excellent starting point to search for stars with similar composition, strict ``chemical tagging,'' i.e., the association of groups of field stars according to their common composition which would suggest a common site of formation \citep{freeman02}, in principle requires a precise knowledge of abundances of several other elements. Nevertheless, it is well known that the behavior of most elements often analyzed in solar-type stars is such that their abundances scale very well with $\feh$ regardless of the place of origin of the star, making them useless in the search for common chemical abundance patterns. A few species, however, are known to show large star-to-star scatter at constant $\feh$, and those should be preferentially employed in the search for groups of stars with a common origin, particularly solar siblings.

\begin{figure*}
\centering
\includegraphics[bb=-115 84 730 707,width=15.5cm]{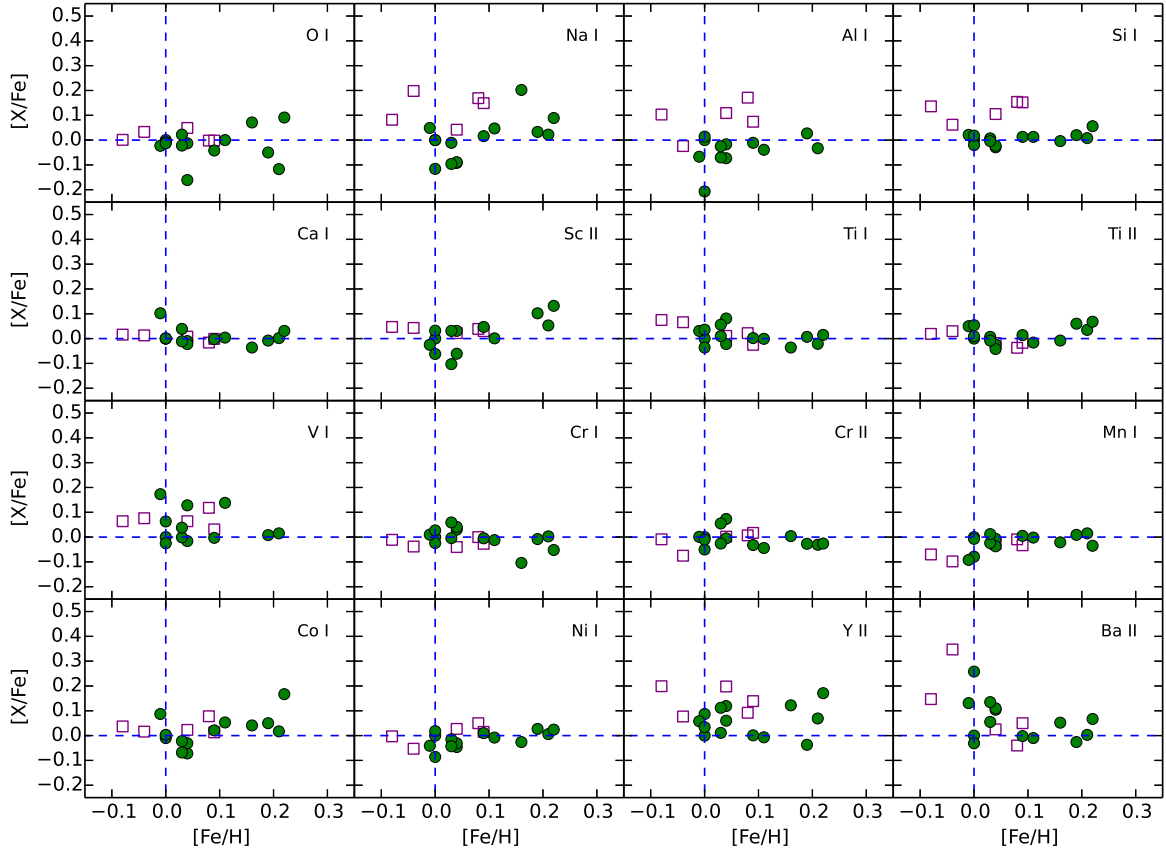}
\caption{Elemental abundance ratios relative to iron as a function of [Fe/H]. Evolved stars are shown with open squares; dwarfs and sub-giants are represented by filled circles. The dashed lines intersect at the solar values.}
\label{f:gce}
\end{figure*}

In \cite{desilva07}, the mean abundance ratios [X/Fe] of a number of elements in fifteen Galactic open clusters were plotted against the mean $\feh$ of each cluster (their Figures 11--15). The $\feh$ range covered by these clusters goes from $-0.6$ to +0.3, but five of them have nearly solar $\feh$ (i.e., $-0.1\lesssim\feh\lesssim+0.1$). With rare outliers, elements like Si, Mg, Ca, Mn, Zr, and Ni have almost indistinguishable [X/Fe] abundance ratios at any given $\feh$. Although some elements such as Ni present very small cluster-to-cluster scatter, the larger scatter in the other cases can be reasonably explained by dissimilar measurement errors (not all elements are equally easy or difficult to analyze). One open cluster of sub-solar metallicity has a very low Zr abundance, but being a single and extreme outlier, that seems to be a very special case and not the rule among these objects. Element Na does seem to have a scatter larger than the expected observational errors, indicating that its abundance is more useful to disentangle stars born in different clusters. A much more obvious case is that of Ba, which for nearly solar-metallicity clusters can vary by almost one order of magnitude. Thus, this study clearly indicates that only Ba, and to some extent Na could be used in practice given the level of uncertainty in these measurements.

\begin{figure*}
\centering
\includegraphics[bb=-115 -29 730 821,width=15.5cm]{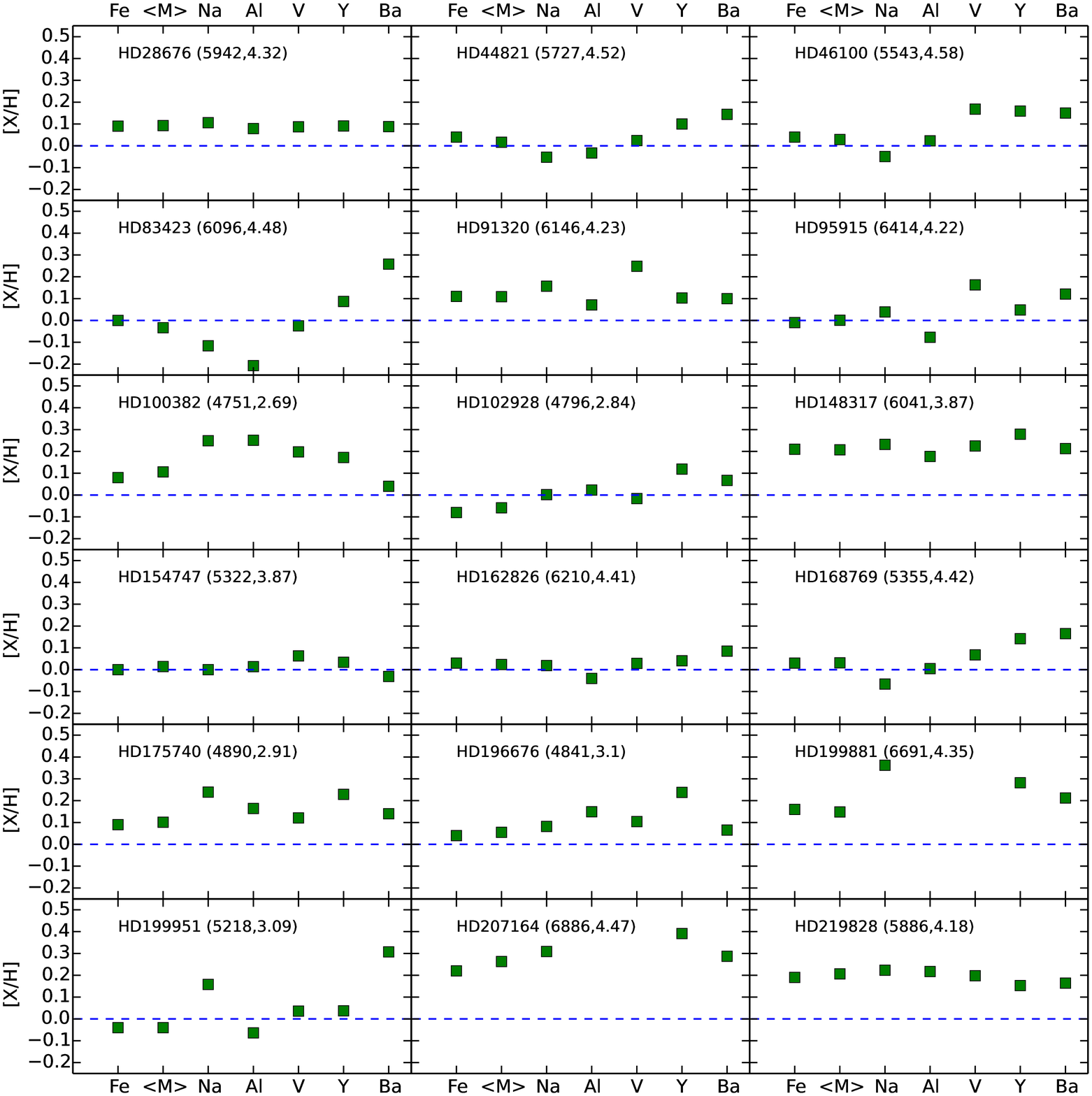}
\caption{Chemical composition of solar sibling candidates. $<\mathrm{M}>$ represents the average of elemental abundances for O, Si, Ca, Sc, Ti, Cr, Mn, Co, and Ni. $\teff$ in K and $\logg$ values are given between parenthesis next to the stars' names.}
\label{f:chemtag}
\end{figure*}

Here we should acknowledge the potential impact of stars with peculiar abundances such as the so-called Ba stars \citep{bidelman51} on our work. These are objects that have prominent \ion{Ba}{2} features in their spectra, which are attributed to close binary interactions, namely mass transfer from an intermediate-mass star that has now evolved into a white dwarf \cite[e.g.,][]{mcclure84,lambert88,han95,husti09}. Quantitatively, their [Ba/Fe] abundance ratios can be as high as 1\,dex at $\feh\simeq0$ \cite[e.g.,][]{smith84,allen06}. This type of metal pollution by a close companion could certainly be problematic for a solar sibling search, although it is possible that the frequency of these events is low enough to have a relatively minor impact. Indeed, the star-to-star scatter of the [Ba/Fe] abundance ratio observed in individual open clusters is low and can be explained purely by observational errors \citep{desilva07}.

Detailed chemical composition studies of large numbers of field stars \cite[e.g.,][]{allende04:s4n,reddy03,takeda07,neves09,gonzalez10,adibekyan12,bensby14} can also guide us in our search. The problem with these large samples, however, is that systematic errors will produce large star-to-star scatter at similar $\feh$ that may prevent us from finding other useful chemical elements. In this context, the work by \cite{ramirez09} is highly relevant. They measured abundance ratios of 20 elements in an important number of so-called solar analog stars. All these objects have spectra very similar to the solar spectrum. Therefore, their atmospheric parameters $\teff$ and $\logg$ are very similar. This means that systematic errors in the chemical analysis can be almost fully removed by employing a strict differential analysis relative to the Sun. Indeed, their [X/Fe] versus $\feh$ trends show very little star-to-star scatter relative to other works using less strict sample selections.

By examining Figure~1 in \cite{ramirez09} it is clear that elements such as Si, Ca, Sc, Ti, V, Cr, Mn, and Ni are useless in the search for solar siblings. The star-to-star scatter of their abundance ratios is fully consistent with the very small observational errors at any given metallicity in the $-0.2<\feh<+0.2$ range. This is in good agreement with the \cite{desilva07} open cluster work. Also consistent with that work is the fact that \cite{ramirez09} find a very large star-to-star scatter for Zn, Y, Zr, and Ba. Thus, those elements are key for our purposes. The few available lines due to Zn and Zr are blended in the majority of our sample stars. Although we are able to measure precise Zn and Zr abundances in the most Sun-like stars in our sample, we excluded these elements from our work in order to maintain homogeneity in the analysis.

The reason $\alpha$- and iron group element abundance ratios [X/Fe] show little or no star-to-star scatter at a given $\feh$ is that they are produced together in supernovae and their production ratios are set by nuclear properties. The differences in abundance ratio scatter with other elements can be understood in terms of the different nucleosynthesis sites and timescales (see \citealt{nomoto13} for a recent overview of Galactic chemical evolution). In particular, in material of roughly solar metallicity, most barium has been produced by s-process nucleosynthesis, which does not occur in the supernovae responsible for the production of the $\alpha$- and iron-group elements. Barium is produced mostly by low-mass AGB (asymptotic giant branch) stars, where the Ba yields are sensitive to the physical conditions at the time of nucleosynthesis and the stellar parameters; Ba production is thus decoupled from Fe production.

\cite{edvardsson93} were among the first to point out a significant correlation between Ba abundance and stellar age at a given $\feh$. Later investigations of open clusters \citep{dorazi09,maiorca11,jacobson13} as well as surveys of field stars \citep{reddy03,bensby07,dasilva12} have confirmed that result, which holds true also for Y.\footnote{Note, however, that \cite{yong12} find only a weak Ba-age correlation in their study of open clusters.} This implies that the scatter in the [Ba/Fe] versus [Fe/H] and [Y/Fe] versus [Fe/H] relations is at least in part due to an age effect. Younger stars tend to exhibit higher barium and yttrium abundances. \cite{maiorca12} have shown that these trends can be reproduced in Galactic chemical evolution models if AGB nucleosynthesis of $M<1.5\,M_\odot$ stars is such that the neutron source is enhanced by a factor of four relative to that of more massive AGB stars.

\subsection{Elemental Abundances of Solar Sibling Candidates}

The abundance ratios relative to iron as a function of $\feh$ for the elements that we measured are shown in Figure~\ref{f:gce}. Evolved stars, in our particular case defined as objects with $\logg<3.5$ (see Figure~\ref{f:hr}) are shown separately from the dwarfs and sub-giants. The reason for this is that it is known that large systematic errors are introduced when comparing these two types of stars. Their extremely different atmospheric structures prevent us from deriving highly accurate abundances for both sets of stars simultaneously. Differential analysis reduces the errors somewhat, but since the reference object (the Sun) is a dwarf, the analysis of giant stars is more suspect. One should therefore be careful when inspecting these trends, because the giant stars may introduce artificial scatter. Indeed, the [Na/Fe], [Al/Fe], [Si/Fe], and [Y/Fe] versus $\feh$ plots clearly show that the giant star sample is shifted upwards with respect to the dwarf stars. Thus, the magnitude of the star-to-star scatter of those elements is amplified by the fact that the analysis of dwarfs and giants is not fully compatible. By examining the [Si/Fe] trends for giants and dwarfs separately, it is clear that the star-to-star scatter is in fact zero within the uncertainties. Therefore, Si is not a useful element in our context, but one could have been misled by the data if dwarfs and giants had not been examined independently.

In good agreement with the open cluster and solar analog studies, we find that most elements present a star-to-star scatter that is fully compatible with the measurement errors. The exceptions are the following species: Na, Al, V, Y, and Ba. Hereafter, all other elements are combined into a single indicator, M.

In Figure~\ref{f:chemtag} we show elemental abundances, relative to H, on a star-by-star basis, separating the important elements in our context (Na, Al, V, Y, and Ba) from M (the combination of all other elements: O, Si, Ca, Sc, Ti, Cr, Mn, Co, and Ni). A star with the same composition as the Sun must have all values in Figure~\ref{f:chemtag} around zero within the errors. Two stars stand out in this context: HD\,154747 and HD\,162826. Both have solar abundances within the errors, although the latter appears to have a slightly super-solar Ba abundance. Another interesting object is HD\,28676, which appears to have a +0.1 offset in the abundances of all elements while retaining almost perfectly solar [X/Fe] abundance ratios. A similar pattern is exhibited by HD\,93210, with the exception of its V abundance that appears very high. The latter, however, could be due to an uncertain effective temperature (see below). From the chemical standpoint, these four objects are key solar sibling candidates.

\subsection{Accounting for Systematic Errors} \label{s:systematics}

In Section~\ref{s:pars} we described our method for deriving atmospheric parameters using only measurements of iron line strength on our high resolution, high signal-to-noise ratio spectra. Stellar properties derived in this manner are often referred to as ``spectroscopic parameters.'' Another common approach to derive the fundamental stellar parameters $\teff$ and $\logg$ involves the use of photometric data (colors) and measured trigonometric parallaxes. The former allow us to constrain $\teff$ from color calibrations based on less model-dependent techniques such as the infrared flux method (IRFM) or even temperatures measured directly from known stellar angular diameters and bolometric fluxes. Parallaxes, on the other hand, allow us to calculate absolute magnitudes of stars, which can then be employed along with theoretical isochrones to compute the stellar masses and thus have another way of estimating $\logg$. The stellar parameters thus derived are sometimes referred to as ``physical parameters.''

In order to assess the impact of systematic errors in our elemental abundance measurements, we re-derived them using physical parameters, which were determined using the procedure outlined in \cite{ramirez13:thin-thick}. Briefly, $\teff$ was measured using as many photometric colors as available and the IRFM $\teff$-color calibrations by \cite{casagrande10}. Surface gravities were then inferred using the stars' {\it Hipparcos} parallaxes and the Yonsei-Yale isochrone grid \citep{yi01,kim02}. All four physical parameters, hereafter referred to as $\teff'$, $\logg'$, $\feh'$, and $\vt'$, were determined iteratively until a final self-consistent solution was achieved, i.e., $\feh'$ and $\vt'$ were re-computed by forcing the iron abundances to be independent of reduced equivalent width, but the excitation and ionization balance conditions were relaxed. Errors in $\teff'$ correspond to the color-to-color scatter, but weighted by the uncertainty of each color-$\teff$ calibration. The error in $\logg'$ was estimated from the width of the isochrone $\logg$ probability distribution (see Section~3.2 in \citealt{ramirez13:thin-thick} for details). Finally, the uncertainty in $\feh'$ was computed by propagating the $\teff'$ and $\logg'$ errors into the iron abundance calculations.

\begin{figure}
\centering
\includegraphics[bb=85 235 538 564,width=8.6cm]{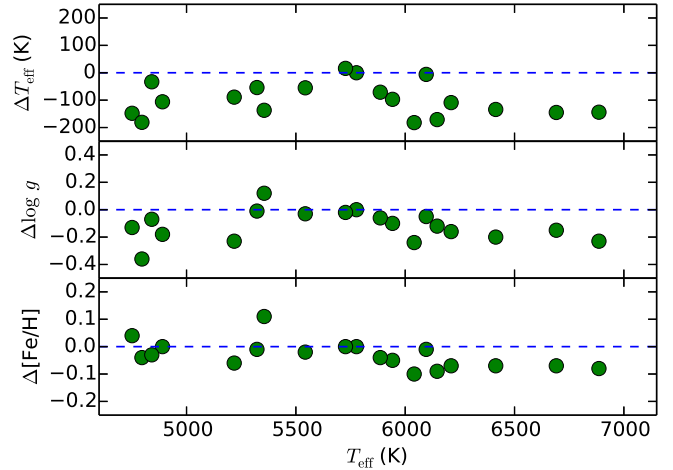}
\caption{Physical minus spectroscopic parameters.}
\label{f:physical}
\end{figure}

Figure~\ref{f:physical} shows the difference between physical and spectroscopic parameters for our sample stars. On average, they are $-97\pm60$\,K for $\teff$, $-0.12\pm0.11$ for $\logg$, and $-0.03\pm0.05$ for $\feh$ (physical minus spectroscopic). Although there are non-negligible systematic offsets in $\teff$ and $\logg$, the $\feh$ values appear relatively robust, particularly for stars with $\teff\lesssim6000$\,K. Depending on the spectral features used, certain elements can be more sensitive to systematic errors in the stellar parameters. Thus, we re-examined Figure~\ref{f:chemtag} for the case of elemental abundances determined using physical parameters. None of the stars previously discarded as solar sibling candidates, i.e., all but the four key targets mentioned in the last paragraph of Section~\ref{s:chemtag}, has its chemical composition affected in such a way that it would resemble the solar abundances had we employed only physical parameters. There are in some cases important variations of the elemental abundances, but generally they are not larger than 0.1\,dex. Thus we conclude that systematic errors in the stellar parameters are important only for those objects which already have near solar abundances (or at least near solar abundance ratios), but are not large enough to force us to re-consider the other targets in our sample as potentially true solar siblings.

\begin{figure*}
\centering
\includegraphics[bb=-115 35 730 763,width=16.0cm]{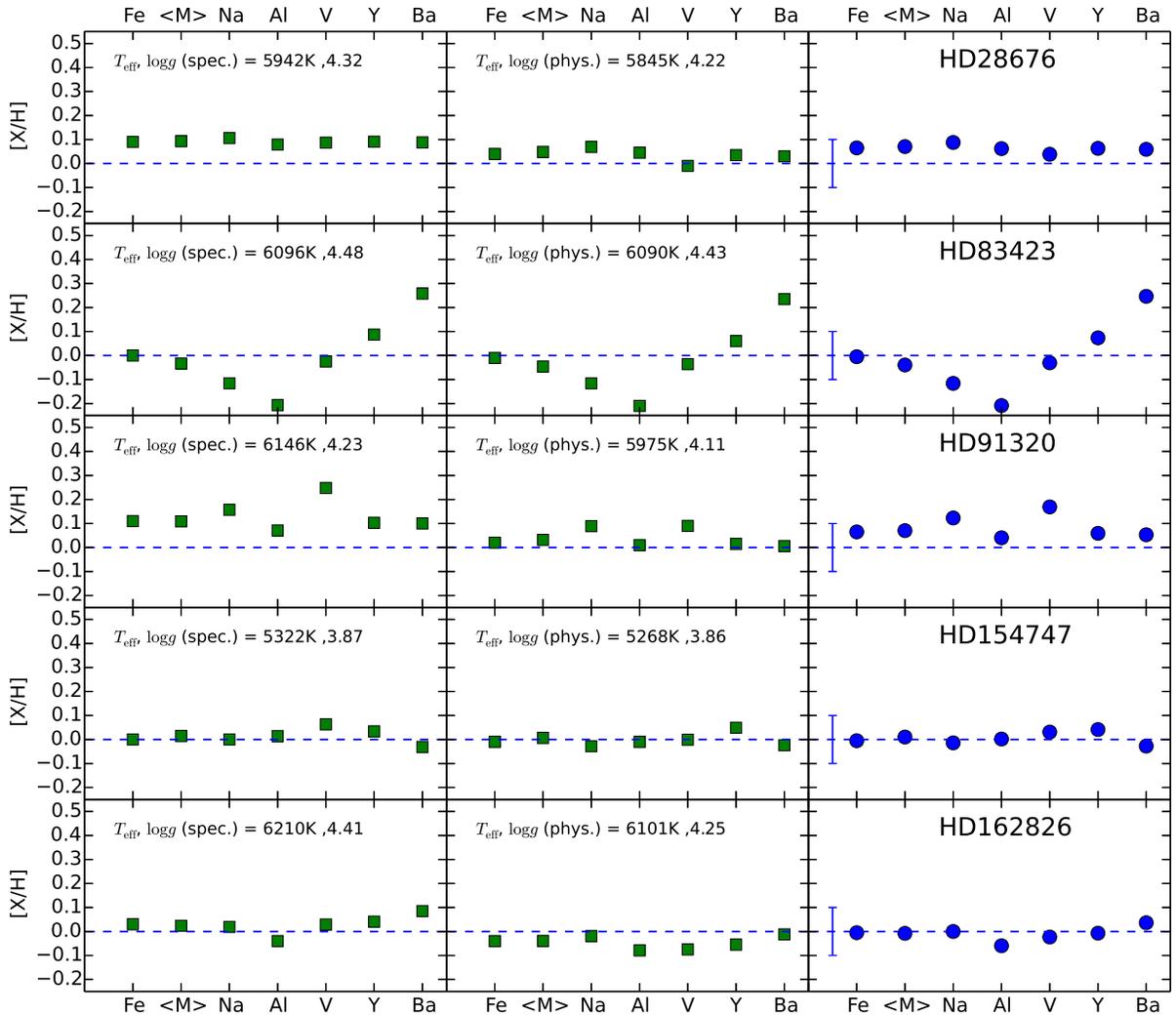}
\caption{Chemical composition of our 5 key solar sibling candidates. Each row corresponds to one star, whose name is provided in the right-most panel. Left panels: as in Figure~\ref{f:chemtag}, i.e., elemental abundances obtained using spectroscopic parameters. Middle panels: as in the left panels, but for elemental abundances derived using physical parameters. Right panels: average of ``spectroscopic'' and ``physical'' abundances. $<\mathrm{M}>$ represents the average of elemental abundances for O, Si, Ca, Sc, Ti, Cr, Mn, Co, and Ni. An error bar of 0.1\,dex, which we estimate as a conservative uncertainty for our abundances, including systematics, is also shown in these panels.}
\label{f:chemtag_5key}
\end{figure*}

In addition to the potential systematic errors introduced by model parameter uncertainties, it is worth mentioning the possibility that the photospheric composition of stars may be affected by planet formation processes. \cite{melendez09:twins} have found that, relative to the majority of solar twin stars, the Sun is deficient in refractory elements by about 0.08\,dex. They attribute this deficiency to the fact that the Sun formed rocky planets, which retained those metals during the formation of the solar system. Similarly, \cite{ramirez11} have found that the secondary star in the 16\,Cyg binary system, which hosts a gas giant planet, is metal-poor relative to the primary, which does not appear to have sub-stellar mass companions \citep{cochran97}. The observed metallicity difference of about 0.04\,dex (volatiles and refractories are equally depleted in this case) is also attributed to the formation of the planet, in this case a gas giant.

It is important to point out that other authors have found conflicting results to the ones described above. In particular, based on an analysis of a stellar sample with a known planet population, \cite{gonzalez-hernandez10,gonzalez-hernandez13} argue that the connection to planet formation processes is weak, although their exoplanet host sample is admittedly biased towards massive planets whereas the \citeauthor{melendez09:twins} hypothesis concerns rocky bodies. In any case, one should keep in mind that the chemical abundance anomalies are still present, and that regardless of their interpretation, they do introduce systematic uncertainties in our context. Similarly, \cite{schuler11:16cyg} have found no differences in chemical composition between the two components of the 16\,Cyg binary system. While this discrepancy with the \citeauthor{ramirez11} results remains unresolved, another study, which employed much higher quality data for these stars and independent measurements of the spectral features, has confirmed the slightly dissimilar chemical composition of the 16\,Cyg stars (Tucci-Maia et~al., in preparation).

From the discussion above, a conservative estimate of our [X/Fe] errors, including model systematics and the potential impact of planet formation on the surface composition of stars, is $\simeq0.1$\,dex. The line-to-line scatter values listed in Tables~\ref{t:abundances_m} and \ref{t:abundances_notm} are not the dominant source of uncertainty for most elements.

\subsection{Key Targets} \label{s:key}

In Section~\ref{s:chemtag} we listed four key targets for siblings of the Sun based on the similarity of their metal abundance ratios to the solar abundances. They are: HD\,28676, HD\,91320, HD\,154747, and HD\,162826. As will be explained in Section \ref{s:dynamics}, HD\,83423 is another interesting candidate, but purely on a dynamical basis. We add this star to our list of key targets to emphasize certain points of our discussion.

The chemical compositions of our 5 key solar sibling candidates are re-examined in Figure~\ref{f:chemtag_5key}. The left panels show the same data plotted in Figure~\ref{f:chemtag}. The middle panels correspond to the elemental abundances derived using physical parameters. The average values of ``spectroscopic'' and ``physical'' chemical composition are shown in the right panels, along with our representative and conservative error bar of 0.1\,dex.

The lower physical $\teff$ values tend to shift the [X/H] abundances downwards, thus making the compositions of HD\,28676 and HD\,91320 more solar, as anticipated. Nevertheless, the former has still a slightly super-solar overall metallicity while the Na and V abundances of the latter still depart from the other elements' nearly constant [X/H] value.

As explained above, HD\,83423 is only included in our list of key targets due to its dynamical properties. Figure~\ref{f:chemtag_5key} clearly shows that the composition of this object is very far from solar, regardless of which set of atmospheric parameters is employed. Note that this is true even though the $\feh$ and [M/H] parameters for this star can be considered fully consistent with the solar values. This finding stresses the importance of identifying and subsequently studying key elements as opposed to simply measuring abundances of ``as many elements as possible'' when it comes to the practical search of stars with common chemical abundance patterns.

The chemical composition of HD\,154747 is solar for both sets of stellar parameters. The only marginally super-solar abundance is that of V, if derived with spectroscopic parameters, but it becomes perfectly solar if physical parameters are employed instead. HD\,162826 has a slightly super-solar overall metallicity if the abundances are derived using spectroscopic parameters, and a slightly sub-solar metallicity for abundances inferred using its physical parameters. Thus, within the systematic (and observational) uncertainties, this star also has a chemical composition that very closely resembles that of the Sun.

Thus, from further examination of chemical abundances and the potential impact of systematic errors, only two of our key targets can be considered real solar sibling candidates: HD\,154747 and HD\,162826.

\subsection{Dynamical Considerations} \label{s:dynamics}

Our target list consists of stars previously suggested by other authors as solar sibling candidates based on their dynamical properties and, in some cases, additional information regarding age and metallicity. The criteria and associated models employed by these authors are somewhat different. Also, the input radial velocities generally differ from those derived in our work using our high-quality spectra. Thus, it is important to re-assess the dynamical properties of our targets in the context of a solar sibling search, in particular those of our five key solar sibling candidates, using a consistent model. In this work, we use the model by \cite{bobylev11}, which is described below. The parallaxes and proper motions of stars were taken from the revised version of the {\it Hipparcos} catalog \citep{vanleeuwen07}.

Stellar and solar orbits were computed using the following Galactic potential:
\begin{equation}
\Phi = \Phi_\mathrm{halo} + \Phi_\mathrm{disk} + \Phi_\mathrm{bulge}
       + \Phi_\mathrm{sp}\ .
\end{equation}

The first three components are constructed as follows \cite[e.g.,][]{helmi04,fellhauer06}: the halo is represented by a logarithmic potential:
\begin{equation}
\Phi_\mathrm{halo} = v_0^2 \ln \left( 1+R^2/d^2 +z^2/d^2 \right)\ ,
\end{equation}
with $v_0=134$\,km\,s$^{-1}$ and $d=12$\,kpc ($R$ and $z$ are cylindrical coordinates); the disk is represented by a Miyamoto-Nagai potential:
\begin{equation}
\Phi_\mathrm{disk} = -\frac
{GM_d}
{\sqrt{R^2+(b+\sqrt{z^2+c^2})^2}}\ ,
\end{equation}
with $M_d=9.3\times10^{10}\,M_\odot$, $b=6.5$\,kpc, and $c=0.26$\,kpc; and the bulge is modeled as a Hemquist potential:
\begin{equation}
\Phi_\mathrm{bulge} = -\frac{GM_b}{r+a}\ ,
\end{equation}
with $M_b=3.4\times10^{10}\,M_\odot$ and $a=0.7$\,kpc.

The following spiral wave component makes this Galactic gravitational potential model more realistic \citep{fernandez08}:
\begin{equation}
\Phi_\mathrm{sp} = A\cos[m(\Omega_p t-\theta)+\chi(R)]\ ,
\end{equation}
where:
\begin{equation}
A = \frac{(R_0\Omega_0)^2 f_{r0}\tan i}{m}\ ,
\end{equation}
and:
\begin{equation}
\chi(R) = -\frac{m}{\tan i}\ln\left(\frac{R}{R_0}\right)+\chi_\odot\ .
\end{equation}

Our adopted spiral wave parameters are: pitch angle $i=-12\degr$, number of arms $m=4$, phase of Sun $\chi_0=-120\degr$, strength $f_\mathrm{r0}=0.05$, and velocity of spiral pattern $\Omega_p=20$\,km\,s$^{-1}$\,kpc$^{-1}$. The circular speed at the solar radius ($R_\odot=8.0$\,kpc) is 220\,km\,s$^{-1}$ and the peculiar solar velocities are ($U_\sun, V_\sun, W_\sun = 10,\,11,\,7$)\,km\,s$^{-1}$
\citep{bobylev10:galpars,schonrich10}.

\begin{figure}
\centering
\includegraphics[bb=110 84 507 707,width=8.2cm]{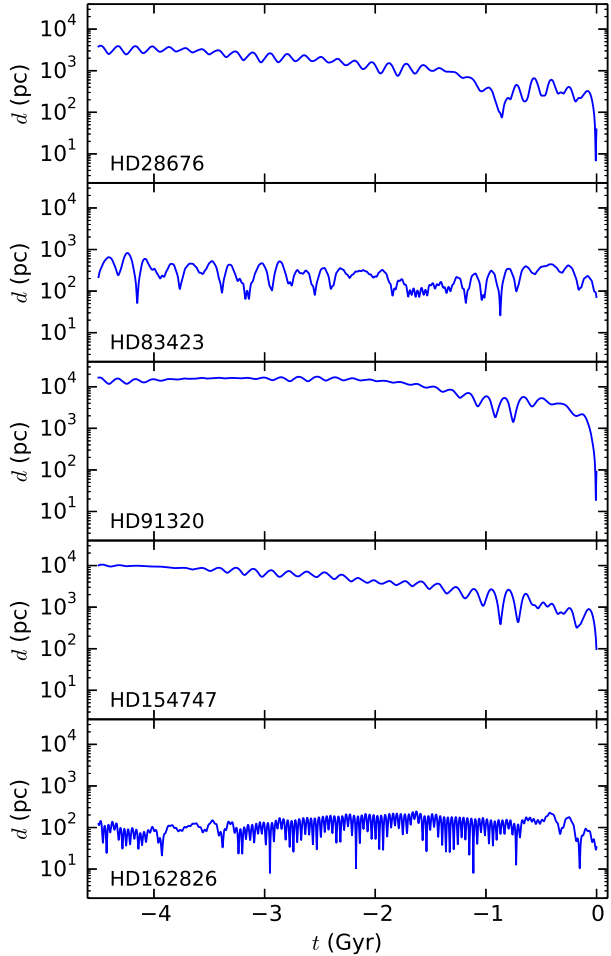}
\caption{Relative distance between the stellar and solar orbits as a function of time in the past.}
\label{f:dynamics_d}
\end{figure}

\begin{figure}
\centering
\includegraphics[bb=110 84 507 707,width=8.2cm]{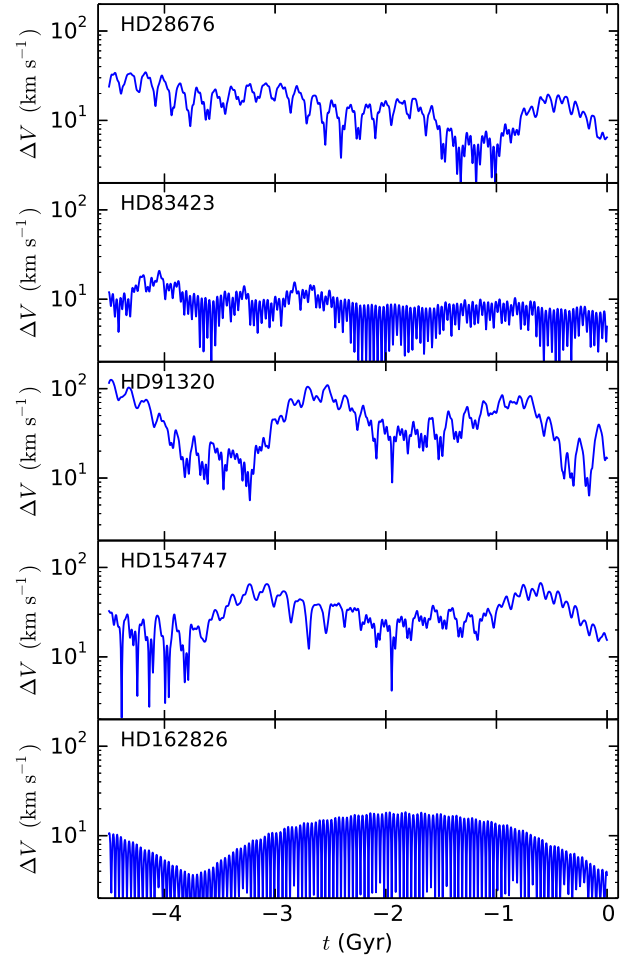}
\caption{Velocity difference between the stellar and solar orbits as a function of time in the past.}
\label{f:dynamics_dv}
\end{figure}

The model described above allows us to compute encounter parameters between the stellar and solar orbit in the past. In particular, we can find the relative distance $d$ and velocity difference $dV$ for the two orbits as a function of time in the past over a 4.5\,Gyr age interval, i.e., the lifetime of the Sun. The results for our five key targets are shown in Figures~\ref{f:dynamics_d} and \ref{f:dynamics_dv}.

A quick inspection of Figures~\ref{f:dynamics_d} and \ref{f:dynamics_dv} clearly reveals that, according to our model, the orbits of HD\,28676, HD\,91320, and HD\,154747 have taken these stars far away from the Sun in the past. In general, the distance between these stars and the Sun has been shortening considerably. If these objects were born also 4.5\,Gyr ago, they may have formed 5--15\,kpc away from the solar cluster. At that time, their velocities relative to the solar motion would have been 30--50\,\kms. Thus, the dynamics rule out these three stars as siblings of the Sun. Of these three objects, only HD\,154747 passed our chemical composition constraint. Not surprisingly, these findings demonstrate that elemental abundance analysis alone is not sufficient, and neither is the dynamical argument by itself. Both are required to make a proper solar sibling identification.

It is also not surprising that of all stars in our sample only HD\,83423 and HD\,162826 passed the dynamical constraint, at least in the sense that their $d$ and $dV$ parameters do not diverge or oscillate with large amplitudes as time recedes. They were the two best solar sibling candidates identified by \cite{bobylev11}, whose model is employed in this work. This shows that relatively small changes to the input radial velocities (we used those measured in our high-resolution, high signal-to-noise spectra, and not those previously published and found in large $RV$ compilations) have a minor impact on these calculations.

As discussed in Section~\ref{s:chemtag}, our chemical analysis quickly ruled out HD\,83423 as a solar composition star. Its $\feh$ value is certainly solar, as is its combined ``M'' abundance (i.e. the average of O, Si, Ca, Sc, Ti, Cr, Mn, Co, and Ni) as well as its V abundance (see Figure~\ref{f:chemtag_5key}). However, the star has a somewhat low (high) Na (Y) abundance, and a definitely too low (high) Al (Ba) abundance. The latter are irreconcilable with the solar values within any acceptable uncertainties in stellar parameters.

This leaves us with only one true solar sibling candidate: HD\,162826. Its chemical composition is solar within the errors and its past orbit includes a number of close encounters ($d\simeq10$\,pc) with the Sun which happened with relative velocities of about 10\,\kms\ or less. The encounter parameters are particularly favorable around $t=-4$\,Gyr, i.e., at an epoch when the solar cluster may have not fully dissipated yet.

We did not consider the impact of interactions with field stars and/or giant molecular clouds in our model. Also, we neglected the gravitational interaction between stars in the solar proto-cluster. According to \cite{garcia01}, the {\it Hipparcos} data suggests a frequency of 2 stellar encounters within one parsec for each $10^6$\,yr (1\,Myr). After correcting for incompleteness, this rate increases to about 12 per Myr.

One should keep in mind, however, that the orbit of a star can change significantly only in very close encounters with massive stars, which are rare and short-lived. The latter leads to a significantly lower rate of encounters compared to M-dwarfs. Indeed, the results from \citeauthor{garcia01} (cf.\ their Table~8) imply that for a period of 4\,Gyr the Sun may have encountered, within 0.1\,pc, only 1 or 2 B-type stars. The theoretical calculations by \cite{ogorodnikov65} are even more reassuring; to change the velocity of a star by more than 20\,\kms\ in an encounter within 40\,AU, the required time period ($\sim10^{13}$\,yr) is longer than the lifetime of our Galaxy. Giant molecular clouds, on the other hand, affect the orbits of the Sun and nearby stars, including any potential siblings, at the same time. Therefore, in this case it is the differential (significantly smaller) impact of such encounters that one would need to take into consideration.

Our dynamical model has a number of input parameters, particularly those related to the spiral arms, which can be varied within reasonable uncertainties to establish the degree of model dependency of our results. Given the demanding nature of these computations, we restricted them to two of our stars: HD\,154747 and HD\,162826, i.e., the two stars that have solar chemical composition. Orbit calculations were made with 2 and 4 arms, varying the pitch angle from $-10$ to $-14$ degrees for the 4-arm model and from $-5$ to $-7$ for the 2-arm model. The phase of the Sun relative to the spiral arm was varied from $-90$ to $-140$ degrees, and the pattern speed was varied from 10 to 24~km\,s$^{-1}$\,kpc$^{-1}$. The total number of models computed is 900 for each star.

For HD\,162826 we obtained past close encounters ($d<100$\,pc, $\Delta V<50$\,km\,s$^{-1}$, $t<-3$\,Gyr) in 64\,\%\ of the models. On the other hand, HD\,154747 shows this type of encounters in only 1.3\,\% of the models. Thus, within reasonable model uncertainty, HD\,162826 remains a good dynamical solar sibling candidate, while it remains highly unlikely that HD\,154747 was born near the Sun.

\begin{deluxetable*}{ccccccccc}
\tablecaption{Rare Earth Element Abundances}
\tablewidth{0pt}
\tablehead{
\colhead{Species} &
\colhead{Wavelength} &
\colhead{EP} &
\colhead{log $gf$} &
\colhead{log $gf$} &
\colhead{hfs/IS} &
\colhead{$\log\epsilon$} &
\colhead{$\log\epsilon$} &
\colhead{$\log\epsilon$} \\
\colhead{} &
\colhead{(\AA)} &
\colhead{(eV)} &
\colhead{} &
\colhead{reference} &
\colhead{reference} &
\colhead{Sun} &
\colhead{HD\,162826\tablenotemark{a}} &
\colhead{HD\,162826\tablenotemark{b}} }
\startdata 
La~\textsc{ii} & 4662.50 & 0.000 & $-$1.24 & 1 & 2 & 1.10 & 1.15 & 1.07 \\
               & 4748.73 & 0.927 & $-$0.54 & 1 &   & 1.16 & 1.26 & 1.17 \\
               & 5303.53 & 0.321 & $-$1.35 & 1 & 2 & 1.12 & 1.25 & 1.14 \\
               & 6390.48 & 0.321 & $-$1.41 & 1 & 2 & 1.18 & 1.23 & 1.13 \\
\hline
Ce~\textsc{ii} & 4486.91 & 0.295 & $-$0.18 & 3 &   & 1.73 & 1.76 & 1.66 \\
               & 4523.08 & 0.516 & $-$0.08 & 3 &   & 1.58 & 1.68 & 1.56 \\
               & 4562.36 & 0.478 & $+$0.21 & 3 &   & 1.61 & 1.68 & 1.58 \\
               & 4628.16 & 0.516 & $+$0.14 & 3 &   & 1.59 & 1.66 & 1.54 \\
               & 5187.46 & 1.212 & $+$0.17 & 3 &   & 1.59 & 1.65 & 1.55 \\
               & 5330.56 & 0.869 & $-$0.40 & 3 &   & 1.67 & 1.72 & 1.62 \\
\hline
Nd~\textsc{ii} & 4446.38 & 0.205 & $-$0.35 & 4 & 5 & 1.32 & 1.44 & 1.32 \\
               & 5234.19 & 0.550 & $-$0.51 & 4 &   & 1.46 & 1.53 & 1.43 \\
               & 5319.81 & 0.550 & $-$0.14 & 4 &   & 1.34 & 1.37 & 1.30 \\
\hline
Sm~\textsc{ii} & 4467.34 & 0.659 & $+$0.15 & 6 & 5 & 0.88 & 1.00 & 0.90 \\
               & 4519.63 & 0.544 & $-$0.35 & 6 &   & 0.94 & 1.07 & 0.96 \\
               & 4537.94 & 0.485 & $-$0.48 & 6 & 5 & 0.97 & 1.09 & 0.97 \\
               & 4676.90 & 0.040 & $-$0.87 & 6 &   & 0.89 & 1.09 & 0.97 
\enddata
\tablerefs{
(1) \citealt{lawler01};
(2) \citealt{ivans06};
(3) \citealt{lawler09};
(4) \citealt{denhartog03};
(5) \citealt{roederer08};
(6) \citealt{lawler06}
}
\tablenotetext{a}{Spectroscopic model parameters}
\tablenotetext{b}{Physical model parameters}
\label{t:ree}
\end{deluxetable*}

\begin{deluxetable*}{ccccccccc}
\tablecaption{Mean Line-by-Line Rare Earth Element Abundance Differences}
\tablewidth{0pt}
\tablehead{
\colhead{Species} &
\colhead{No.\ lines} &
\multicolumn{3}{c}{HD\,162826\tablenotemark{a} $-$ Sun} &
\colhead{} &
\multicolumn{3}{c}{HD\,162826\tablenotemark{b} $-$ Sun} \\
\cline{3-5} \cline{7-9}
\colhead{} &
\colhead{} &
\colhead{$\langle\Delta\rangle$} &
\colhead{$\sigma$} &
\colhead{$\sigma_{\mu}$} &
\colhead{} &
\colhead{$\langle\Delta\rangle$} &
\colhead{$\sigma$} &
\colhead{$\sigma_{\mu}$} }
\startdata 
La~\textsc{ii} & 4 & $+$0.083 & 0.039 & 0.020 & & $-$0.013 & 0.033 & 0.017 \\
Ce~\textsc{ii} & 6 & $+$0.063 & 0.023 & 0.010 & & $-$0.043 & 0.018 & 0.007 \\
Nd~\textsc{ii} & 3 & $+$0.073 & 0.045 & 0.026 & & $-$0.023 & 0.021 & 0.012 \\
Sm~\textsc{ii} & 4 & $+$0.143 & 0.039 & 0.019 & & $+$0.030 & 0.035 & 0.017 
\enddata
\tablenotetext{a}{Spectroscopic model parameters}
\tablenotetext{b}{Physical model parameters}
\label{t:ree2}
\end{deluxetable*}

\subsection{Abundances of Trace Elements in HD\,162826}

We use several of the rare earth elements to further test whether HD\,162826 meets our chemical criteria for being a solar sibling. In solar-type stars, these elements owe their origin to both $r$-process and $s$-process neutron-capture reactions, which reflect a different set of chemical evolution clocks than the lighter elements. We identify 17~lines of 4~species that are unblended in our spectra of the Sun and HD\,162826. These lines are listed in Table~\ref{t:ree}. Abundances of La~\textsc{ii}, Ce~\textsc{ii}, Nd~\textsc{ii}, and Sm~\textsc{ii} are derived from spectrum synthesis using MOOG. Linelists were constructed using the \citet{kurucz95} lists, using updated $\log gf$ values from recent laboratory studies when possible. We adjust the line strengths to reproduce the solar spectrum and then use these lists without change for the analysis of HD\,162826. Table~\ref{t:ree} lists the wavelength, excitation potential (EP), and $\log gf$ value for each transition, although the transition probabilities cancel out in a differential analysis. Our syntheses account for hyperfine splitting (hfs) and isotope shifts (IS) for several of these lines. We adopt the solar isotopic fractions given by \citet{lodders03} for the Sun and HD\,162826. The derived abundances are listed in Table~\ref{t:ree}. Table~\ref{t:ree2} lists the mean line-by-line differential abundances between the Sun and HD\,162826. Two sets of values are given, one using the set of spectroscopic model atmosphere parameters for HD\,162826 and one using the set of physical values.

Using spectroscopic parameters, the La, Ce, and Nd are very similar to the Ba abundance, i.e., slightly supersolar ($\simeq+0.08$). However, using physical parameters, all these elements have solar abundances in HD\,162826 within the internal error. The Sm abundance is also solar within the errors if we employ the physical parameters, but super-solar at +0.14 dex using spectroscopic parameters. The average of these abundances (excluding Sm), as derived using both sets of parameters, is about +0.02\,dex, i.e., solar within both systematic and internal errors. Of all elements studied in this work for HD\,162826, only Sm appears to depart from the solar abundances, with a spectroscopic/physical average value of $+0.09$. However, considering our conservative estimate of 0.1\,dex of systematic error, this value may be marginally consistent with the solar Sm abundance.

\subsection{High-precision Radial Velocity Data for HD\,162826}

\begin{figure}
\centering
\includegraphics[bb=115 265 509 536,width=7.6cm]{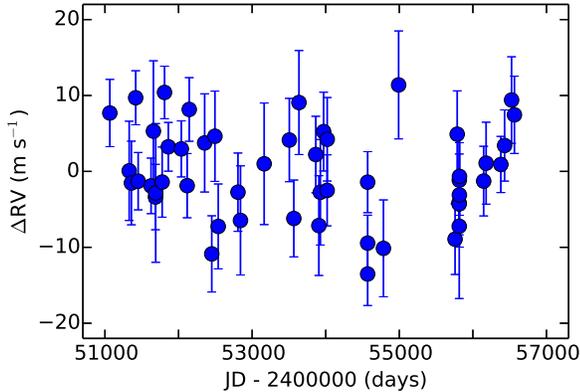}
\caption{Relative radial velocity as a function of Julian Date for HD\,162826. The radial velocities in this plot are given with respect to the weighted mean of all observed values.}
\label{f:hr6669_rvs}
\end{figure}

HD\,162826 is included in the target sample of 250 F, G, K, and M-type stars of the McDonald Observatory planet search program at the Harlan J.\ Smith 2.7\,m Telescope \cite[e.g.,][]{cochran97,endl04,robertson12}. This long-term radial velocity (RV) survey is designed to probe the population of gas giant planets beyond the ice line at several AU. Such planets presumably have not migrated inwards from the location of their formation. Figure\,\ref{f:hr6669_rvs} displays the 15 years of precise RV measurements of HD\,162826. The 50 RV data points have an overall rms-scatter of 6.0\,m\,s$^{-1}$ and an average error of 5.4\,m\,s$^{-1}$. The star is constant at the 6\,m\,s$^{-1}$ level and does not seem to have a massive planetary companion with a period of $<15$ years.  Also, there is no clear evidence of binarity.

\begin{figure}
\centering
\includegraphics[bb=115 265 509 536,width=7.6cm]{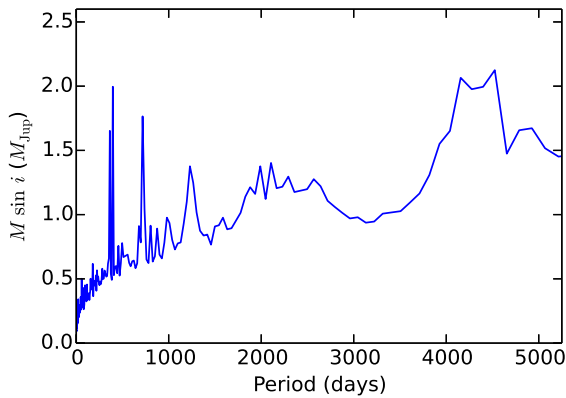}
\caption{Mass limits for single planets in circular orbits around HD\,162826. Planets with parameters in the region above the solid line would have been recovered with 99\,\%\ probability at a false-alarm-probability of less than 1\,\%.}
\label{f:hr6669_limits}
\end{figure}

We computed the upper limits on detectable planets in the RV data for HD\,162826. The detection limit was determined by adding a fictitious Keplerian signal to the data, then attempting to recover it via a generalized Lomb-Scargle periodogram \citep{zechmeister09}. Here, we have assumed circular orbits; for each combination of period $P$ and RV semi-amplitude $K$, we tried 30 values of orbital phase. A planet is deemed detectable if 99\% of orbital configurations at a given $P$ and $K$ are recovered with a false-alarm probability \citep{sturrock10} of less than 1\%.  This approach is essentially identical to that used in the work by \cite{wittenmyer06,wittenmyer10,wittenmyer11:jupiters}. The resulting mass limits are shown in Figure~\ref{f:hr6669_limits}. Clearly, hot Jupiters, i.e., planets with masses comparable to that of Jupiter in short-period orbits, can be ruled out.

To determine the probability that an undetected Jupiter-mass planet orbits HD\,162826, we repeated the detectability simulations described above for a range of recovery rates (10\%...90\%) as in \citet{wittenmyer11:earths}.  For a Jupiter-mass planet in a Jupiter-like (12~yr) circular orbit, we estimate a 35\,\% probability that such a planet is present based on the non-detection from our RV data.

\section{CONCLUSIONS}

Detailed elemental abundance analysis and proper chemical tagging are both required in the search for the stars that were born together with the Sun. However, one should keep in mind that not all elements are equally important. Although deriving abundances of ``as many elements as possible'' would be ideal, in practice one could concentrate on a few key elements as an intermediate step between employing only photometric metallicities and a very detailed high-precision chemical analysis. In particular, the spectral lines due to Ba are very strong, hence easily measured in medium resolution spectra. Moreover, the large star-to-star dispersion observed in the [Ba/Fe] versus [Fe/H] plane suggests that the Ba abundance is highly sensitive to the place of origin of stars. A perfectly reasonable intermediate step would therefore be the measurement of Fe and Ba abundances in medium resolution, moderately high signal-to-noise ratio spectra.

Only the star HD\,162826 (HR\,6669, HIP\,87382) satisfies both our dynamical and chemical criteria for being a true sibling of the Sun. This object, a late F-type dwarf star located at about 34\,pc ($\sim110$\,light years) from the Sun in the constellation Hercules, is bright ($V=6.7$) and easily observable with small and medium-sized telescopes. High-precision radial velocity observations carried out over a period of time longer than 15\,years rule out the presence of hot-Jupiter planets. These data also suggest a 2/3 chance that a Jupiter analog is not present either. Smaller terrestrial planets cannot be ruled out at this moment.

The mass of HD\,162826, estimated from the location of the star on the HR diagram compared to Yonsei-Yale isochrones (as in \citealt{ramirez13:thin-thick}), is $1.15\,M_\odot$. If this star were the only solar sibling in the 1.1 to $1.2\,M_\odot$ range, the \cite{salpeter55} initial mass function would suggest the existence of another $\sim400$ solar siblings of mass greater than $0.1\,M_\odot$ dispersed throughout the Galaxy. Since the number of stars in the solar cluster is estimated to be $10^3-10^4$, this implies that there should be just a few other solar siblings of $\sim1\,M_\odot$ present in the solar neighborhood. On the other hand, this means that there are a few hundred M-dwarfs that are also siblings of the Sun within 100\,pc. Unfortunately, the detailed chemical composition analysis of M-dwarfs that would be required to identify them is currently beyond our capabilities.

The combination of astrometric data from the ongoing Gaia Mission and spectroscopic data from surveys of comparable large size such as the ESO-Gaia survey, APOGEE, and/or GALAH will allow us to discover many more solar siblings in a very near future. We expect that the analysis presented in this paper will guide future endeavors in this field and allow us to perform these searches more efficiently.

\acknowledgments

{\small
I.R.\ acknowledges support from NASA's Sagan Fellowship Program to conduct most of the observations presented in this paper. A.T.B and V.V.B.'s work is supported by the ``Nonstationary Phenomena in Objects of the Universe'' Programme of the Presidium of the Russian Academy of Sciences and the ``Multiwavelength Astrophysical Research'' grant no.\ NSh-16245.2012.2 from the President of the Russian Federation. D.L.L.\ thanks the Robert A.\ Welch Foundation of Houston, Texas for support through grant F-634. The McDonald Observatory exoplanet survey is currently supported by the National Science Foundation under grant AST \#1313075. Previous support for this project was given by the National Aeronautics and Space Agency through the Origins of Solar Systems Program grants NNX07AL70G, NNX09AB30G and NNX10AL60G.
}

\end{document}